\documentclass[11pt,a4paper]{article}

\usepackage{fullpage}

\usepackage{booktabs} 

\usepackage{latexsym}
\usepackage{url}
\usepackage{xcolor}
\usepackage{longtable}
\usepackage{multirow}

\usepackage{xspace}

\usepackage{listings}
\definecolor{mygreen}{rgb}{0,0.5,0}
\definecolor{mygray}{rgb}{0.5,0.5,0.5}
\definecolor{mymauve}{rgb}{0.58,0,0.82}
\usepackage{cite}

\newtheorem{propose}{Proposition}

\newcommand{\JLT}[1]{}
\newcommand{\MP}[1]{}

\usepackage{tikz}
\usetikzlibrary{arrows.meta}
\tikzset{>={Classical TikZ Rightarrow[width=1.7mm,length=1mm]}}

\graphicspath{{./graphics/benchmarks/}}

\newcommand{\eg}{e.g., }
\newcommand{\ie}{i.e., }

\newcommand{\prevnode}{\emph{prev}\xspace}
\newcommand{\nextnode}{\emph{next}\xspace}
\newcommand{\lastnode}{\emph{last}\xspace}

\lstMakeShortInline[keywordstyle=\color{black},basicstyle=\fontsize{8}{8}\selectfont\ttfamily]|

\begin{document}

\title{Stamp-it: A more Thread-efficient, Concurrent Memory Reclamation
  Scheme in the C++ Memory Model}

\author{
Manuel P\"oter \\
TU Wien, Faculty of Informatics\\
Vienna, Austria\\
\url{manuel@manuel-poeter.at}
\and 
Jesper Larsson Tr\"aff\\
TU Wien, Faculty of Informatics\\
Vienna, Austria\\
\url{traff@par.tuwien.ac.at}
}

\maketitle

\begin{abstract}
We present \emph{Stamp-it}, a new, concurrent, lock-less memory
reclamation scheme with amortized, constant-time (thread-count
independent) reclamation overhead. Stamp-it has been implemented and
proved correct in the C++ memory model using as weak
memory-consistency assumptions as possible. We have likewise
(re)im\-ple\-men\-ted six other comparable reclamation schemes. We
give a detailed performance comparison, showing that Stamp-it performs
favorably (sometimes better, at least as good as) than most of these
other schemes while being able to reclaim free memory nodes earlier.
\end{abstract}

\section{Introduction}

Efficient, dynamic memory management is at the heart of many
sequential and parallel algorithms, consisting in the allocation of
pieces of memory and the subsequent, \emph{safe} reclamation of these
pieces when they are no longer in use.  In parallel and concurrent,
lock- and wait-free algorithms, the reclamation step is highly
non-trivial since more than one \emph{thread} may be referencing and
using an allocated piece of memory unbeknownst to other threads: It
cannot be given back to the system or thread-local heap before it has
been ascertained that no threads will possibly access any data in this
memory anymore.

There has been a substantial amount of work on memory reclamation for
concurrent algorithms, see, 
\eg~\cite{Alistarh:2014:StackTrack,Alistarh:2015:TAS:2755573.2755600,Balmau:2016:FRM:2935764.2935790,Braginsky:2013:Drop,Brown:2015:RML:2767386.2767436,Cohen:2015:AMR:2814270.2814298,Cohen:2015:EMM:2755573.2755579,DiceHerlihyKogan16,Gidenstam:2009,Herlihy:2002:ROP,RamalheteC17,Michael:HazardPointers,Hart:2007:PMR:1316099.1316427,Detlefs01lock-freereference,Sundell:2005,WenIzraelevitz18}.
All of these schemes have their merits and (performance) issues. One
drawback shared by them all, except for reference counting schemes, is
that they need to scan references from \emph{all threads} in order to
reclaim possibly no longer referenced memory pieces. A main motivation
of this work is to overcome this bound.

Our contribution is a new lock-less reclamation scheme, called
\emph{Stamp-it}, which is compared qualitatively and experimentally to
six well-known and, depending on circumstances, well performing
current schemes. Reclamation in Stamp-it is done in \emph{amortized
  constant time per reclaimed memory block}; no references are scanned
unless they can be reclaimed.  All tested schemes have been
(re)implemented in C++; full source code is available at
\url{http://github.com/mpoeter/emr}.  The experimental evaluation is
done on four architecturally different systems with large numbers of
hardware supported threads, ranging from 48 up to 512. We use standard
benchmarks, as well as a new benchmark designed to study memory
consumption by reclaimable but unreclaimed memory.  On these
benchmarks and machines, Stamp-it compares favorably to and in many
cases and aspects significantly outperforms the competing schemes.

In the following, a contiguous piece of memory allocated from the
system heap for use in a concurrent algorithm and possibly shared
between threads is called a \emph{node}. We do not deal with memory
management (allocation and deallocation of nodes) here.  Nodes may
store additional meta-information that is not visible to the
application, and we mention where such is required.  

We are interested in \emph{general purpose} reclamation schemes that
allow eventually reclaimed nodes to be freely reused at a later time,
regardless of how and in which data structure the allocated nodes were
used. Not all reclamation schemes have this property,
\eg~\cite{Valois:Phd,Sundell:2005} do not allow general reuse of
reclaimed nodes, \cite{Braginsky:2013:Drop,Gidenstam:2009} have to be
tailored to the application data structure
and~\cite{Cohen:2015:EMM:2755573.2755579,Cohen:2015:AMR:2814270.2814298}
require the data structure to be in a special, normalized form.  A
general purpose scheme should be \emph{non-intrusive}, requiring no or
little changes in the application code. A way of achieving this is to
rely on a standard interface as proposed for
C++~\cite{Robison:2013}. A reclamation scheme should be fast, both in
use and maintenance of references to shared nodes, as well as in the
actual reclamation. It should require little memory overhead, avoid
typical performance issues like \emph{false sharing} and should not
prevent applications using commonly found patterns in lock-free
programming like borrowing some bits from a pointer.  Reclaimability
of nodes should be detected fast to reduce unnecessary memory
consumption.  Robustness against crashes, and bounds on the amount of
memory blocked by crashed threads are desirable but not provided by
most schemes. We say that a reclamation scheme is \emph{lock-less},
if, provided that sufficient system memory is available, no thread can
block the progress of the application. On the other hand, a scheme is
\emph{reclamation-blocking} if a suspended or crashed thread can
prevent an unbounded amount of nodes from being
reclaimed~\cite{Hart:2007:PMR:1316099.1316427}.  With this
terminology, Stamp-it is lock-less but reclamation-blocking.
Lock-freedom should allow good scalability; wait-freedom would be
desirable, but not many schemes actually provide this.

All lock- and wait-free algorithms rely on hardware supported atomic
operations. We consider only solutions that use standard atomics
like \emph{fetch-and-add} (FAA) and \emph{single-word
  compare-and-swap} (CAS). Solutions requiring non-standard
double-word compare-and-swap (as in,
\eg~\cite{Detlefs01lock-freereference}) will either be non-portable or
require expensive emulations. We also rule out solutions that have to
be tailored to specific data structures
like~\cite{Braginsky:2013:Drop, Gidenstam:2009}, or require hardware
or operating system specific features like transactional memory,
\eg~\cite{Alistarh:2014:StackTrack} or POSIX signals,
\eg~\cite{Alistarh:2015:TAS:2755573.2755600,Brown:2015:RML:2767386.2767436}.
The aim was to design a portable, fully C++ standard conform and
platform independent implementation. Our implementation is mature
beyond a simple proof of concept, and applicable for real-life
applications. It works with \emph{arbitrary numbers of threads} that
can be started and stopped arbitrarily.

Based on the above discussion, we have implemented six comparable
schemes that make the same assumptions. These are Lock-free Reference
Counting (LFRC)~\cite{Valois:Phd}, Hazard Pointers
(HP)~\cite{Michael:HazardPointers}, Quiescent State-based Reclamation
(QSR)~\cite{Mckenney98read-copyupdate}, Epoch-based Reclamation
(ER)~\cite{Fraser:Phd}, New Epoch-based Reclamation
(NER)~\cite{Hart:2007:PMR:1316099.1316427}, and
DEBRA~\cite{Brown:2015:RML:2767386.2767436}. Interval-based
Reclamation (IR) would fit among these, but is too recent to be
considered~\cite{WenIzraelevitz18}. Hart et
al.~\cite{Hart:2007:PMR:1316099.1316427} used a similar selection of
schemes in their study, and we wanted to repeat their experiments with
our own implementations on different platforms and at a larger
scale. Common to all implementations, including Stamp-it, is that we
rely on the C++ memory model and can argue that our implementations
are correct in this model. We try to use as weak consistency
assumptions as possible.

In the remainder of this report we first illustrate how a reclamation
scheme for C++ can be used in lock-free algorithms and data structures
with the C++ interface that we support~\cite{Robison:2013}. In
Section~\ref{sec:stampit} we describe our new reclamation scheme
Stamp-it and argue for correctness and amortized complexity.  An
experimental comparison between Stamp-it and the other six schemes is
given in Section~\ref{sec:experiments}. More results can be found in
the appendix, a previous technical
report~\cite{Traff17:reclamationcorr,Traff18:reclamation} and
in~\cite{Poeter18} which is the basis for this work.

\section{C++ Memory Reclamation Interface}
\label{sec:interface}

Before presenting the reclamation schemes, we illustrate how memory
reclamation can be done in actual applications using an interface
proposed for C++ by Robison~\cite{Robison:2013}.  This proposal
defines an abstract interface and allows many different reclamation
schemes to be implemented and used. The interface defines the
following fundamental pointer abstractions:
\begin{itemize}
\item A |marked_ptr| allows one or more low-order bits to be borrowed.
  Many lock-free algorithms rely on such mark tricks,
  \eg~\cite{Harris:2001:PIN:645958.676105,Boehm:2004:ANS:1011767.1011774,SundellTsigas:2008}.  The |get|
  method returns the raw pointer (without the mark bits), and the
  |mark| method returns the value of the mark bits.
\item A |concurrent_ptr| acts like an atomic |marked_ptr|, \ie it
  supports atomic operations.
\item A |guard_ptr| is an object that can atomically take a snapshot
  of the value of a |concurrent_ptr| and guarantee that the target will
  not be deleted as long as the |guard_ptr| holds a pointer to it.
\end{itemize}
It is important to note that only |guard_ptr| references protect
against deletion of a node. In effect, a |concurrent_ptr| is a
``weak'' pointer and a |guard_ptr| is a ``shared ownership'' pointer.

To obtain a snapshot from a |concurrent_ptr| and populate a
|guard_ptr|, |acquire| and |acquire_if_equal| methods can be used. In
wait-free algorithms, |acquire| may be problematic with some schemes
like Hazard Pointers, or even LFRC, because it may have to loop
indefinitely. For such cases, |acquire_if_equal| can be used as it
simply stops trying if the value in the pointer does not match an
expected value, and reports whether it was successful or not.

Releasing a |guard_ptr| follows the standard smart pointer
interface. For a |guard_ptr| instance |g|, the operation |g.reset|
releases ownership and sets |g| to |nullptr|; the |guard_ptr|
destructor implicitly calls |reset|.  The |reclaim| method marks the
given node to be reclaimed once it is safe to do so and implicitly
resets the |guard_ptr|.

An implementation of the |find| function in Harris' list-based
set~\cite{Harris:2001:PIN:645958.676105} with the improvements
proposed by Michael~\cite{Michael:2002:HPD:564870.564881} using this
interface is shown in Listing~\ref{lst:list:find}.

\begin{lstlisting}[caption={Implementation of list::find}, label=lst:list:find]
template <class Key, class Reclaimer>
bool list<Key, Reclaimer>::find(Key key,
  concurrent_ptr*& prev, marked_ptr& next,
  guard_ptr& cur, guard_ptr& save) {
retry:
  prev = &head;
  next = prev->load();
  save.reset();
  for (;;) {
    if (!cur.acquire_if_equal(*prev, next)) 
    goto retry;
    if (!cur) return false;
    next = cur->next.load();
    if (next.mark() != 0) {
      next = cur->next.load().get();
      marked_ptr expected = cur.get();
      if (!prev->compare_exchange_weak(
            expected, next)) goto retry;
      cur.reclaim();
    } else {
      if (prev->load() != cur.get()) goto retry;
      Key ckey = cur->key;
      if (ckey >= key) return ckey == key;
      prev = &cur->next;
      save = std::move(cur);
    }
  }
}
\end{lstlisting}
We first acquire a |guard_ptr| to the next node and store it in |cur|,
ensuring that we can safely iterate the list. We use the
|acquire_if_equal| method since we already know the expected value. We
then check if |cur|'s |next| pointer has the mark bit set. In that
case we try to splice out the node, and mark |cur| for reclamation.
Otherwise, we continue to iterate the list. We \emph{move} |cur| into
|save|, \ie the previous |guard_ptr| in |save| (if any) gets reset and
replaced with the value from |cur|; and |cur| is reset.

We made a number of small changes and adaptations to Robisons
interface proposal~\cite{Robison:2013}. These changes are described
in detail in~\cite{Poeter18}. The most important change is
the introduction of the |region_guard| concept, which is required for
reclamation schemes like NER, QSR and Stamp-it. In these schemes a
|guard_ptr| can only exist inside a \emph{critical region}, so unless
the thread is already inside a critical region the |guard_ptr|
automatically enters one. Entering and leaving critical regions are
usually rather expensive operations, and |region_guard|s allow to
amortize this overhead. Any |guard_ptr| instances created inside the
scope of a |region_guard| can simply use the current critical region
and save the overhead of entering a new one.  In QSR, the
|region_guard|s are used to reduce the number of \emph{fuzzy barriers}.

\section{Stamp-it Memory Reclamation}
\label{sec:stampit}

We now introduce our new scheme, \emph{Stamp-it}. It is an
\emph{epoch-based scheme} conceptually similar to NER and
therefore provides many of the same properties. As in ER/NER, the
programmer has to define \emph{critical regions} that are entered and
left explicitly. A thread is only allowed to access shared objects
inside such regions. 

The Stamp-it algorithm maintains thread-local and one global
\emph{retire-list} of nodes that can potentially be reclaimed, and
relies on an abstract data structure called \emph{Stamp Pool} that
efficiently supports the following operations:
\begin{enumerate}
    \item Add an element and assign a stamp to it (|push|). Stamps
      have to be strictly increasing, but not necessarily consecutive.
    \item Remove a specific element, return |true| if this element was
      the one with the lowest stamp at that point in time (|remove|).
    \item Get the highest stamp assigned to an element so far, \ie the
      last stamp that has been assigned to an element.
    \item Get the lowest stamp of all elements currently in the Stamp Pool.
\end{enumerate}
The algorithm uses the Stamp Pool as follows. Upon entering a critical
region the thread adds itself to the pool, and gets a new stamp
value. Stamp values are strictly increasing, therefore, defining a total
order in which all threads have entered their respective critical region.
When a thread retires a node it requests the highest stamp from the Stamp
Pool, stores it in the node and appends the node to the end
of its local retire-list.

\begin{figure}[tbh]
\centering
\begin{tikzpicture}[xscale=6.5,yscale=1.0, font=\footnotesize]
    \draw[->] (-0.05,1.4) node[above]{Threads} -- (-0.05,0) -- (1.05,0) node[below] {$time$};
    
    \node [align=left, right] at (-.22, -0.9) {\texttt{head} stamp};
    \node [align=left, right] at (-.22, -1.25) {\texttt{tail} stamp};
    \node [align=left, right] at (-.22, -1.8) {Thread stamps};
    
    \draw               (-0.05,  0.4) -- (0.95, 0.4) node[right] {$T_1$};
    \draw [ultra thick] (0.1,  0.4) -- (0.8, 0.4);
    
    \draw               (-0.05, 0.8) -- (1.0, 0.8) node[right] {$T_2$};
    \draw [ultra thick] (0.25, 0.8) -- (0.67, 0.8);
    
    \draw               (-0.05, 1.2) -- (1.0, 1.2) node[right] {$T_3$};
    \draw [ultra thick] (0.5, 1.2) -- (0.94, 1.2);
    
    \node []          at (.02, -0.9) {0};
    \node []          at (.02, -1.25) {0};
    
    \draw [dotted, mygreen] (0.1,  0.4) -- (0.1, -0.2) node[below] {$t_1$};
    \draw [dotted, mygreen, ->] (0.1, -0.55) -- (0.1, -0.75) node[right,font=\fontsize{5}{5}\selectfont] {update};
    \node [mygreen]          at (0.1, -0.9) {1};
    \node [align=left,font=\fontsize{7}{7}\selectfont] at (0.1,  -1.8) {$s_1$=0\\$s_2$=-\\$s_3$=-};
    
    \draw [dotted, mygreen]     (0.25,  0.8)  -- (0.25, -0.2) node[below] {$t_2$};
    \draw [dotted, mygreen, ->] (0.25, -0.55) -- (0.25, -0.75) node[right,font=\fontsize{5}{5}\selectfont] {update};
    \node [mygreen]          at (0.25, -0.9) {2};
    \node [align=left,font=\fontsize{7}{7}\selectfont] at (0.25,  -1.8) {$s_1$=0\\$s_2$=1\\$s_3$=-};
    
    \node [above, blue]   at (0.4,  0.4) {remove $n_1$};
    \draw [dotted, blue]     (0.4,  0.4)  -- (0.4, -0.2) node[below] {$t_3$};
    \draw [dotted, blue, <-] (0.4, -0.55) -- (0.4, -0.75) node[right,font=\fontsize{5}{5}\selectfont] {read};
    \node [blue]          at (0.4, -0.9) {2};
    
    \draw [dotted, mygreen]     (0.5,  1.2)  -- (0.5, -0.2) node[below] {$t_4$};
    \draw [dotted, mygreen, ->] (0.5, -0.55) -- (0.5, -0.75) node[right,font=\fontsize{5}{5}\selectfont] {update};
    \node [mygreen]       at (0.5, -0.9) {3};
    \node [align=left,font=\fontsize{7}{7}\selectfont] at (0.5,  -1.8) {$s_1$=0\\$s_2$=1\\$s_3$=2};
    
    \node [above, blue]   at (0.61,  0.8) {remove $n_2$};
    \draw [dotted, blue]     (0.61,  0.8)  -- (0.61, -0.2) node[below] {$t_5$};
    \draw [dotted, blue, <-] (0.61, -0.55) -- (0.61, -0.75) node[right,font=\fontsize{5}{5}\selectfont] {read};
    \node [blue]          at (0.61, -0.9) {3};
    
    \draw [dotted, red]     (0.67,  0.8) -- (0.67, -0.2) node[below] {$t_6$};
    \node [align=left,font=\fontsize{7}{7}\selectfont] at (0.67,  -1.8) {$s_1$=0\\$s_2$=-\\$s_3$=2};
    
    \draw [dotted, red]     (0.8,  0.4) -- (0.8, -0.2) node[below] {$t_7$};
    \node [align=left,font=\fontsize{7}{7}\selectfont] at (0.8,  -1.8) {$s_1$=-\\$s_2$=-\\$s_3$=2};
    \node [red]          at (0.8, -1.25) {2};
    \draw [dotted, red, ->] (0.8, -0.55) -- (0.8, -1.1) node[right,font=\fontsize{5}{5}\selectfont] {update};
    
    \draw [dotted, red]     (0.94, 1.2) -- (0.94, -0.2) node[below] {$t_8$};
    \node [align=left,font=\fontsize{7}{7}\selectfont] at (0.94,  -1.8) {$s_1$=-\\$s_2$=-\\$s_3$=-};
    \node [red]          at (0.94, -1.25) {3};
    \draw [dotted, red, ->] (0.94, -0.55) -- (0.94, -1.1) node[right,font=\fontsize{5}{5}\selectfont] {update};
\end{tikzpicture}
\caption{An example sequence of Stamp Pool operations. 
Thick lines indicate critical regions.}
\label{fig:stamp-it-concept}
\end{figure}
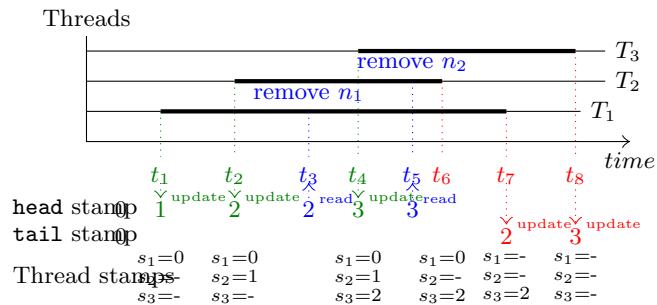

Upon leaving a critical region the thread removes itself from the Stamp
Pool. and performs a reclaim operation on the local retire-list.
The reclaim operation requests the lowest stamp from the Stamp Pool.
It can then safely reclaim all nodes with a smaller stamp
value. Since nodes are appended to the end, the elements in the local
retire-list are ordered by their stamp values. This makes reclamation
very efficient as it always has a runtime linear in the number of
nodes that can currently be reclaimed; no time is wasted on nodes that
cannot yet be reclaimed.
The working of Stamp-it is illustrated in Figure~\ref{fig:stamp-it-concept}.

If the |remove| operation returns |false| and the number of nodes in the
local retire-list exceeds some threshold, the thread pushes all
remaining entries to the global retire-list as an ordered sublist. If
|remove| returns |true|, \ie the thread was the last one,
it performs a reclaim operation on the global retire-list. The
global retire-list is not ordered and therefore does not provide the
same runtime guarantees. However, since it is organized as a list of
ordered sublists, each sublist needs to be scanned only up to the node
which has a stamp that is larger than or equal to the lowest stamp.
Therefore, the resulting total runtime is $O(n+m)$ where
$n$ is the total number of reclaimable nodes and $m$ is the number of
ordered sublists.

We will argue for the following propositions, and deal with progress
properties later:
\begin{propose}
Stamp-it is \emph{reclamation safe}: A node is reclaimed only when it is
referenced by no thread.
\end{propose}
Assume that some thread retires (want to reclaim) some node
$n$. It fetches the currently highest stamp from the Stamp Pool,
stores this stamp value in $n$ and adds $n$ to its local retire-list.
Now, $n$ can safely be reclaimed once all threads that were in a critical
region at the time $n$ was removed have left their respective critical
regions. The lowest stamp in the Stamp Pool is smaller than or equal
to the stamp of the last thread, so $n$'s stamp being less than this lowest
stamp implies that all threads currently inside critical regions (if any)
have entered their respective critical region \emph{after} $n$ was
retired, and therefore $n$ can safely be reclaimed (no references).

\begin{propose}
Stamp-it reclaims any node in amortized constant time in the number of
Stamp Pool operations.
\end{propose}
Assume that $q$ threads are inside critical regions, when thread $T_1$
is leaving its critical region and thus removing itself from the Stamp
Pool. All nodes in $T_1$'s local retire-list can be reclaimed
once these $q$ threads have left their critical regions. Thread $T_1$
can possibly add its local retire-list as sublist to the global
retire-list, deferring reclamation of its nodes to the \emph{last}
thread to leave its critical region.  The global retire-list is
traversed only when this last thread removes itself from the Stamp
Pool. In the worst case, the $q$ threads remove themselves in the
same order as they have entered their critical regions, resulting in
$q$ traversals of the global retire-list. We can therefore achieve
amortized constant time by only adding local retire-lists that hold
more than $q$ nodes (threshold); the sublist might be touched up to
$q$ times to reclaim at least $q$ nodes.

Maintaining a dynamic counter for $q$ can cause additional overhead.
Alternatively, one could use a less volatile counter as upper bound
such as the total number of threads, but then the total number of
nodes left in local retire-lists would be quadratic in the number of
threads and would therefore increase unnecessary memory
consumption. For this reason, we use a \emph{static threshold} with an
empirical value of 20 in our implementation.

Our implementation of the Stamp Pool is built on the ideas of the
lock-free doubly-linked list by Sundell and
Tsigas~\cite{SundellTsigas:2008}. It requires two static dummy nodes,
|head| and |tail|, which are also used to manage the highest
and lowest stamp values; the highest stamp is stored in |head|
and the lowest one in |tail|.  

Our implementation differs from~\cite{SundellTsigas:2008} in that we
only need to push a node from one direction (next to |head|), and that
each node can be removed at any time, independent of its
position. Each thread has a single local |thread_control_block| that
acts as a node in the Stamp Pool, \ie the nodes are ``reused'' and
we therefore have to take care of the ABA problem and consider the
possibility that nodes might reappear at different positions. We will
from now on refer to the |thread_control_block|s in the Stamp Pool
as \emph{blocks} (not to be confused with reclaimable memory nodes).

To insert or delete a block from the Stamp Pool one has to update
the respective set of |prev| and |next| pointers. These have to be
changed consistently, but not necessarily all at once. The solution
proposed by Sundell and Tsigas is to treat the doubly-linked list as a
singly-linked list with auxiliary information in the |prev| pointers.
Thus, the |next| pointers always form a consistent singly-linked list,
but the |prev| pointers only give hints for where to find the previous
block. We use the same approach, but reversed the directions, \ie keep
the |prev| list consistent and use the |next| pointers as auxiliary
information, as this is more suited to the use in Stamp-it.

Both pointers, |next| and |prev| have to be equipped with a deletion
mark (in the least significant bit) to prevent conflicting updates
from concurrent insert and delete operations as in Harris'
singly-linked list~\cite{Harris:2001:PIN:645958.676105}.

To avoid the ABA problem, in addition to the delete mark we spare
additional 17 bits for a \emph{version tag} in both pointers. These bits
are used to store a tag that gets incremented with every change to the
pointer value. There is still a very small chance for an undetected ABA to
occur when the version tag wraps around, but in order for this to happen
there have to be \emph{exactly} $2^{17}$ updates to the pointer between the
initial read and the subsequent CAS operation.

\subsection{Stamp Pool data structure}

The Stamp Pool keeps track which threads have entered a critical
region and in which order. It holds two static dummy blocks |head| and
|tail|; new blocks are inserted right after |head|.  The |prev|
pointers define the direction from |head| to |tail|; this direction is
always kept consistent. The |next| pointers define the direction from
|tail| to |head|, but only act as hints where to find the next
block. It is therefore possible that a block, which is already
in the |prev| list, does not occur (yet) in the |next| list (and
the other way round in case of removal).

Each queue block, including |head| and |tail|, holds a \emph{stamp counter}.
When a new block is inserted, it loads |head|'s |prev|
pointer and stores it in its local |prev|, increases |head|'s
|stamp| using an FAA operation, stores the returned value in its local
|stamp| and then performs a CAS on |head->prev| in order to insert
itself into the |prev| list. This ensures that |head| always holds the
highest stamp and that the stamps in the |prev| direction are strictly
decreasing, \ie the stamp of the newly added block is greater than all
other blocks (except |head|).  The only exception to this is the
|tail| block which should always reflect the stamp value of
its immediate predecessor in the |prev| direction.  This way we can easily
fetch the lowest stamp value from |tail|.

Even though the |next| pointers only act as hints, it is guaranteed
that they only point to blocks with a higher stamp value, \ie the stamps
in the |next| direction are strictly increasing (with the |tail| block again
being an exception). More specifically, the |next| pointer of some block
$b$ can point to:
\begin{itemize}
    \item |head|: This can be the case when |head| is the
      predecessor of $b$, or when some other block $c$ has inserted
      itself between |head| and $b$, but did not yet update $b$'s
      |next| pointer. Note that when $b$ has already marked its |prev|
      pointer, it can no longer be updated by $c$, so this
      inconsistency can only be resolved once $b$ is fully removed
      from both lists.
    \item a block $c$ which is still in the |prev| list: This is the
      ``normal case'', \ie usually $c$ is the predecessor of $b$;
      unless $b$ is |head|, in which case the previously described
      exception is possible.
    \item a block $c$ which is removed from the |prev| list: This is
      the intermediate state when $c$ has been removed from the |prev|
      list, but not yet from the |next| list. So |prev->next| points
      to $c$, but $c$ is no longer the immediate predecessor of $b$ in
      the |prev| direction when starting from |head|. However, by
      following $c$'s |next| pointer (and potentially those of other
      removed blocks) one can find $b$'s new predecessor. 
\end{itemize}

\begin{figure}[!htb]
\begin{center}
\begin{tikzpicture}[xscale=0.7, yscale=0.62, font=\footnotesize]
\node at ( 0,  0) [rectangle,draw] (tail) {tail};
\node at ( 1.25,1.2) [rectangle,draw] (t1) {$T_1$};
\node at ( 2.5,  0) [rectangle,draw] (t2)   {$T_2$};
\node at ( 3.75,1.2) [rectangle,draw, color=red] (t3)   {$T_3$};
\node at ( 5,0) [rectangle,draw] (t4)   {$T_4$};
\node at ( 7,0) [rectangle,draw] (t5)   {$T_5$};
\node at ( 9,0) [rectangle,draw,color=green!60!black] (t6) {$T_6$};
\node at ( 11,0) [rectangle,draw] (head) {head};

\path[->, font=\fontsize{6}{6}\selectfont]
	[->] (head) edge [bend right=-18] node [below] {prev} (t6)
	[->] (t6)   edge [bend right=-25] node [below] {prev} (t5)
	[->] (t5)   edge [bend right=-25] node [below] {prev} (t4)
	[->] (t4)   edge [bend right=-25, color=red] node [below] {prev} (t2)
	[->] (t3)   edge [bend right= 20, color=red] node [above, sloped] {prev} (t2)
	[->] (t2)   edge [bend right=-25] node [below] {prev} (tail)
	[->] (t1)   edge [bend right=-20, color=red] node [below, sloped] {prev} (tail)

	[->] (tail) edge [bend right=-20] node [above, sloped] {next} (t1)
	[->] (t1)   edge [bend right=-20, color=red] node [above, sloped] {next} (t2)
	[->] (t2)   edge [bend right=-25] node [above] {next} (t4)
	[->] (t3)   edge [bend right=-20, color=red] node [above, sloped] {next} (t4)
	[->] (t4)   edge [bend right=-45, color=red] node [above] {next} (head)
	[->] (t5)   edge [bend right=-35] node [above] {next} (head)
	[->] (t6)   edge [bend right=-15] node [above] {next} (head);
\end{tikzpicture}
\end{center}
\caption{Example of blocks with their links; green blocks are not yet fully
         inserted, red blocks are removed; red links are marked.}
\label{fig:stamp-it-queue}
\end{figure}
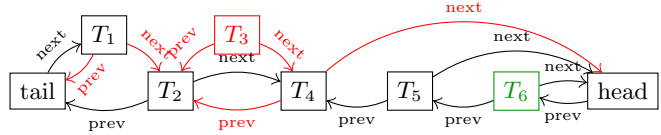
An example is shown in Figure~\ref{fig:stamp-it-queue}.
The links of the blocks $T_1$, $T_3$ and $T_4$ are all marked so they cannot
be updated. The block $T_3$ is already fully removed, \ie it is not referenced
by any |prev| nor |next| pointer in the list. The blocks $T_1$ and $T_4$ are
marked for deletion, but are not yet fully removed; $T_1$ has been removed from
the |prev| list, but is still in the |next| list; $T_4$ is still fully linked.
Block $T_5$ already finished its |push| operation (not
green), but the |next| pointer of its successor, $T_4$, still points
to |head|. This indicates that $T_4$'s |next| pointer was already
marked, so it could not be updated by $T_5$. On the other hand, $T_6$
is currently in the |push| operation (green); it has successfully inserted
itself in the |prev| list, but the update of $T_5$'s |next| pointer is
still pending.

The two lowest bits of the stamp counter are used to embed flags to
track a block's state:
\begin{itemize}
    \item |PendingPush|: The block is currently being inserted into
      the |prev| list.
    \item |NotInList|: The block has been completely removed, \ie 
      is no longer part of neither |prev| nor |next| list. This
      implies that the owning thread is no longer inside a critical
      region.
\end{itemize}
The two flags are mutually exclusive, so they cannot both be set at
the same time. A block can be in four different states:
\begin{itemize}
    \item \emph{in the process of being inserted into the prev list}:
      The |PendingPush| flag is set, the |NotInList| flag and |next|'s
      delete mark are cleared and |prev|'s delete mark is
      undetermined.
    \item \emph{in the queue}: The delete marks of the
      |prev| and |next| pointer, as well as the |NotInList| flag are
      cleared, the |PendingPush| flag is undetermined. Note that this
      state only indicates that the block was correctly inserted in the
      \emph{prev} direction, but does not state anything about the
      \emph{next} direction since we cannot update the |next| pointer
      of a block that is already marked.
    \item \emph{in the process of being removed}: The delete mark of
      the |prev| pointer is set, the mark of |next| is undetermined
      and the |NotInList| and |PendingPush| flags are cleared.
    \item \emph{fully removed}: The |NotInList| flag and
      |prev|'s delete mark are set, the |PendingPush| flag is cleared
      and |next|'s delete mark is undetermined (because it gets reset
      in the |push| operation before the |PendingPush| flag is set).
\end{itemize}

\subsection{The Basic Steps in Detail}

We now show how to support the operations on the doubly-linked list
data structure correctly in a lock-free manner. Some code listings
are included in the main text, the rest can be found in the appendix.
For brevity, we omitted the C++ memory orderings from the code shown
here. Full code listings with memory orderings and more detailed
explanations are provided in~\cite{Poeter18}.

To insert a block we first set the block's |next| pointer to |head|
implicitly clearing the |DeleteMark| of |next|. Then we perform an FAA
on |head->stamp|, getting the new stamp for the block we are about to
insert. This new stamp value is modified to have the |PendingPush|
flag set before it is stored in our block. Then we set the block's
|prev| pointer and attempt a CAS operation to update |head->prev| with
our own block.  The implementation of the |push| method is shown in
Listing~\ref{lst:push} in the appendix.  When the CAS is successful
our block has been inserted in the |prev| list. We can therefore reset
the |PendingPush| flag, and perform a final CAS-loop to update our
successor's |next| pointer.

When a thread leaves a critical region, it removes itself from the
queue.  The |remove| operation first marks the |prev| and |next|
before removing the block from both lists. Then it sets the
|NotInList| flag and checks if this thread was the last one, \ie if
the block's |prev| pointer points to |tail|. If that is the case, it
tries to update |tail|'s |stamp| to that of the new ``last'' thread.

Marking the two pointers signals to other threads that this block is
about to be removed, and also prevents the pointers from being updated
by CAS operations from threads that did not yet see the mark. In order
to remove a block $b$ from the |prev| list, the thread has to find its
predecessor, \ie the block $c$ with the |prev| pointer pointing to
$b$, and update $c$'s |prev| pointer with the value of $b$'s |prev|
pointer. But it can of course happen that $c$'s |prev| pointer is also
marked and can therefore not be updated. In this case we have to find
$c$'s predecessor and \emph{help} remove $c$ before we can continue
with the removal of $b$. By removing $c$, we get a new predecessor for
$b$. We can then restart the loop and try to remove it again. The same
idea is applied when removing a block from the |next| list.

Since a block $b$ can only be removed from the |prev| list when its
immediate predecessor is not marked, any marked immediate predecessor
has to be removed before $b$ can be removed. Therefore, whenever a
thread that tries to remove a marked block $b$ encounters another
block $c$ which is supposed to come \emph{after} $b$ in the |prev|
direction (\ie it was found by following the |prev| pointers starting
from $b$), where $\mathrm{stamp}_c > \mathrm{stamp}_b$, or
$\mathrm{stamp}_c$ has the |NotInList| flag set, and all blocks on the
path from $b$ to $c$ are also marked, we can conclude that $b$ has
already been removed from both lists.

Since $c$ was encountered \emph{after} $b$ in the |prev| direction, it
is supposed to have a lower stamp than $b$; it can only have a larger
stamp if it was removed and then reinserted. But since all blocks
between $c$ and $b$ are marked, $c$ could not have been removed
without first removing all those blocks, including $b$. The same holds
for the case when the |NotInList| flag is set, as the flag is only set
once the block has been fully removed.

The code to remove a block from the |prev| list is shown in
Listing~\ref{lst:remove-from-prev}.
The remove operation keeps track of three different pointers:
\begin{description}
    \item[\emph{prev}] is a reference to the next \emph{unmarked}
      block in the |prev| direction, \ie the block that we want to set as
      the new value for our predecessor's |prev| pointer. We get to
      this block by following our own |prev| pointer and the |prev|
      pointers of other marked blocks (if any).
    \item[\emph{next}] is a reference to some block that precedes
      our own block in the |prev| direction. By following this block's
      |prev| pointer we should end up at our own block, unless some
      other thread has removed it already. This way we can efficiently
      find our immediate predecessor to update its |prev| pointer.
    \item[\emph{last}] is a reference to a helper block that is used
      to remove potentially marked predecessors of our own block.
      When this pointer is not null, it should be the immediate
      predecessor of the \nextnode block in the |prev| direction.
\end{description}
So the order of the blocks in the |prev| direction should be as follows:
\lastnode (if it is set), \nextnode, our own block $b$, and \prevnode. Each
of these blocks (except $b$) can potentially be removed and reinserted at any
time. For \nextnode and \lastnode we have to consider this possibility and
take appropriate actions. However, when we recognize that \prevnode has
been removed or reinserted, we can stop since we know that $b$ must
have been removed already as well.

\begin{lstlisting}[caption={Stamp-it's remove\_from\_prev\_list method}, label=lst:remove-from-prev]
bool remove_from_prev_list(
  marked_ptr& prev, marked_ptr b, marked_ptr& next)
{
  const auto my_stamp = b->stamp.load();
  marked_ptr last = nullptr;
  for (;;) {
    if (next.get() == prev.get()) {
      next = b->next.load(); return false;
    }
    auto prev_prev = prev->prev.load();
    auto prev_stamp = prev->stamp.load();
    if (prev_stamp>my_stamp || prev_stamp&NotInList)
      return true;
    if (prev_prev.mark() & DeleteMark) {
      if (!mark_next(prev, prev_stamp)) return true;
      prev = prev->prev.load(); continue;
    }
    auto next_prev = next->prev.load();
    auto next_stamp = next->stamp.load();
    if (next_prev != next->prev.load()) continue;
    if (next_stamp < my_stamp) {
      next = b->next.load(); return false;
    }
    if (next_stamp & (NotInList | PendingPush)) {
      if (last.get() != nullptr) {
        next = last; last.reset();
      } else next = next->next.load();
      continue;
    }
    if (remove_or_skip_marked_block(
          next, last, next_prev, next_stamp))
      continue;
    if (next_prev.get() != b.get()) {
      move_next(next_prev, next, last); continue;
    }
    if (next->prev.CAS(next_prev, prev))return false;
  }
}
\end{lstlisting}
The |remove_from_prev_list| operation essentially consists of a large
loop that keeps track of the three mentioned blocks, while trying to
find the direct predecessor of $b$. Once that predecessor is found, we
can try to update its |prev| pointer in order to remove $b$. There are
several conditions that lead to the termination of this loop. In some
of these cases we can conclude that $b$ is already removed from
\emph{both} lists, in other cases we know that $b$ has been removed
from the |prev| list, but we still need to ensure that it is also
removed from the |next| list.

In line 7, if \prevnode and \nextnode point to the same block, $b$ must have
been removed from the |prev| list already. If in line 12 the \prevnode's
stamp is greater than $b$'s |stamp| or has the |NotInList| flag set,
we can conclude that \prevnode must have been removed (together with
$b$).

If \prevnode's |prev| pointer is marked (line 14), we try to help setting
the delete mark on \prevnode's |next|. The |mark_next| operation simply
performs a CAS loop trying to set the delete mark on the |next|
pointer as long as that block's |stamp| matches the given stamp
value. When we detect that the |stamp| has changed, we can conclude
that \prevnode must have been removed (together with $b$).

If \nextnode's |stamp| is less than $b$'s stamp (line 21), we can
conclude that $b$ must have been removed from the |prev|
list. Otherwise, if \nextnode's stamp has the |NotInList| or
|PendingPush| flag set (line 24) we cannot use this block since it
might not be part of the |prev| list.  For the |NotInList| flag this
is clear, but for the |PendingPush| flag this is 
more subtle: The flag signals that the block is currently getting
inserted into the |prev| list, but with the information we have
available at this time it is impossible to tell whether this has
already happened or not.

In line 30 we call the helper function |remove_or_skip_marked_block|
(shown in Listing~\ref{lst:remove-or-skip-marked-block} in
the appendix). This method checks whether the \nextnode block is marked,
and if so, tries to remove it provided we have a valid
\lastnode pointer (recall that \lastnode is supposed to be the predecessor
of \nextnode). In case we have no \lastnode pointer, we move \nextnode to the
next block in the |next| direction.

If \nextnode's |prev| pointer does not match $b$ (line 33) it is not
$b$'s predecessor. In that case we call |move_next| (shown in
Listing~\ref{lst:move-next}) and restart the loop.

The |move_next| method tries to move \nextnode to the following block in
the |prev| direction, while keeping the old value of \nextnode in
\lastnode. There is a special case that needs to be handled. It could
happen that the next block in the |prev| direction has successfully
inserted itself into the list, but still has the |PendingPush| flag
set, \ie it did not yet finish its |push| operation. We previously
checked that \nextnode's |stamp| does not have the |PendingPush| flag
set, because otherwise we would have to dismiss the block as we could
not determine whether it is already inserted. Now we can conclude that
it is in fact part of the |prev| list, so we help resetting the
|PendingPush| flag. This is necessary to ensure lock-freedom, as
otherwise we would iterate infinitely because in the next iteration
the previously mentioned check would fail again, resulting in \nextnode
being moved back in the |next| direction again.
\begin{lstlisting}[caption={Stamp-it's move\_next method},
  label=lst:move-next]
void move_next(
    marked_ptr next_prev, marked_ptr& next, marked_ptr& last)
{
  size_t next_prev_stamp = next_prev->stamp.load();
  if (next_prev_stamp & PendingPush &&
      next_prev == next->prev.load()) {
    auto expected = next_prev_stamp;
    auto stamp=next_prev_stamp+StampInc-PendingPush;
    if (!next_prev->stamp.CAS(expected, stamp))
      if (expected != stamp) return;
  }
  last = next; next = next_prev;
}
\end{lstlisting}
When we arrive at line 37 we have found $b$'s predecessor and can
attempt a CAS on \nextnode's |prev| with our current \prevnode to remove $b$
from the |prev| list.

In case we could not conclude that $b$ is already removed from both
lists, we still have to remove it from the |next| list; this is done
in the |remove_from_next_list| operation (shown in
Listing~\ref{lst:remove-from-next} in the appendix), which is quite
similar to |remove_from_prev_list|. It also keeps track of the same
three pointers, where the initial values for \prevnode and \nextnode are
those that were returned by |remove_from_prev_list|.  This allows us
to continue from where we left, reducing the amount of work to find
the blocks we need to update in many cases.

In this method we have to set \nextnode to the last unmarked block with a
stamp greater than $b$'s |stamp|, and \prevnode to the first unmarked
block with a stamp less or equal to $b$'s |stamp| (both in the |prev|
direction), \ie the two blocks that would be the predecessor and
successor of $b$ if $b$ would still be part of the |prev| list. This
entails that \nextnode's |prev| pointer must reference \prevnode. Once we
have found these blocks, we can attempt a CAS to update \prevnode's
|next| pointer in order to finish removal of $b$ from the |next|
list. If the CAS succeeds, we have successfully removed $b$. However,
we still have to make sure that the \prevnode block has not been marked
in the meantime. If this is the case, we have to continue and help
remove \prevnode from both lists in order to maintain the previously
described condition, which lets us conclude that a block has been
fully removed if we recognized that the successor block has been fully
removed.

The |update_tail_stamp| method (shown in Listing~\ref{lst:update-tail}
in the appendix) tries to find the new predecessor of |tail| in the |prev|
direction, read its stamp and update |tail|'s stamp
accordingly. Unfortunately, finding this predecessor is not as simple
as taking |tail|'s next pointer, since it could point to |head| (due
to the predecessor not having finished its push operation) or to a
block that could have been removed and potentially reinserted at the
time we read its |stamp|. Of course we could detect such cases and try
to find the actual predecessor, but we do not want to waste too much
time for this. Instead, if we cannot immediately identify the new
predecessor we simply use the ``next best guess'' for the new stamp,
which is our own block's |stamp| plus a stamp-increment (recall that
stamps are strictly increasing).

Finally, we perform a simple CAS-loop, trying to update |tail|'s
|stamp| as long as the new value we want to write is greater than the
value we are trying to replace.

\subsection{Correctness}

We can now argue for the following proposition:

\begin{propose}
Stamp-it is \emph{lock-less}, that is, all methods used for entering
and leaving critical regions (|push|, |remove|), and all helper functions
are lock-free. The expected average runtime of the operations is constant, if
threads do not conflict.
\end{propose}
Stamp-it is, however, reclamation-blocking in the sense that a stalled (or
crashed) thread can prevent an unbounded number of nodes for being
reclaimed~\cite{Hart:2007:PMR:1316099.1316427}.

The |push| operation (see Listing~\ref{lst:push} in the appendix) is
lock-free. The first loop performs a CAS operation in order to insert
the block into the queue. In case the CAS succeeds, we break out of
the loop, otherwise we just restart the loop. The CAS can only fail if
some other thread interfered---either by inserting or removing some
block. But in this case some other thread must have made progress.
The same argument applies to |update_tail_stamp| and
|mark_next|. Both methods contain loops that perform CAS operations,
and a failure of these operations can only be caused by progress in
some other thread.

The |remove_from_prev_list| (see Listing~\ref{lst:remove-from-prev})
and |remove_from_next_list| (see Listing~\ref{lst:remove-from-next} in
the appendix) operations are a bit more complex. We argue for
|remove_from_prev_list|, and since both operations are similar, the
same will hold for |remove_from_next_list|. As mentioned, both keep
track of a \prevnode and a \nextnode pointer. In each iteration we perform
one of the following changes in case we have to restart the loop:
\begin{itemize}
	\item move \prevnode along the |prev| direction (in case \prevnode
          is marked)
	\item move \nextnode along the |prev| direction (in case \nextnode
          is not \prevnode's predecessor)
	\item remove \nextnode from the |prev| list (in case \nextnode is
          marked and we have a valid \lastnode pointer)
	\item move \nextnode along the |next| direction (in case \nextnode
          is marked and we have no \lastnode pointer, or \nextnode has the
          |NotInList| or |PendingPush| flag set)
	\item nothing (in case the CAS to remove $b$ failed)
\end{itemize}
The block $b$ divides the |prev| list into two sublists: The sublist
from |head| to $b$, and the sublist from $b$ to |tail|. \nextnode points
to a block in the first sublist and \prevnode points to a block in the
second sublist. New blocks are inserted at the beginning of the |prev|
list (right after head), \ie they become part of the first sublist. So
the number of times we can move \nextnode in the |prev| direction before we
reach $b$ is bounded by the number of entries in the first sublist,
and the number of times we can move \prevnode in the |prev| direction before
we reach |tail| is bounded by the number of entries in the second
sublist.

The case where we have to move \nextnode back in the |next| direction
because it is marked and we have no valid \lastnode can be resolved by
following \nextnode's |next| pointer and from there move again along
the |prev| pointer, while maintaining \lastnode. So the next time we
encounter the same marked block, we will be able to remove it as we
now have a valid \lastnode pointer. In the worst case scenario we have
to move along the |next| direction until \nextnode points to |head|,
from where we can then start to move \nextnode along the |prev|
direction, while removing marked blocks (if any). The case where
\nextnode has the |PendingPush| flag set can be resolved in the same
way.

This leaves us with the cases where \nextnode has the |NotInList| flag
set or the CAS operation to remove $b$ fails. But both cases can only
occur when another thread changed the data structure in a way that
it is no longer consistent with the thread's view. So unless some
other thread interferes, for both methods, |remove_from_prev_list| and
|remove_from_next_list|, it is guaranteed that at any time a thread is
able to finish the method in a bounded number of steps.

Unfortunately, the block pointed to by \nextnode can be removed and
reinserted at any time. Obviously, this destroys the bounds as with
every reinsertion the block is put back right at the beginning of the
|prev| list. However, this implies that the owning thread of this
reinserted block has been able to finish its |remove| and subsequent
|push| operation, \ie that it has made progress. Thus, the requirements for
lock-freedom are fulfilled as it is guaranteed that at any time at
least one thread makes progress: If there is no conflict with another
thread, we can finish the operation in a bounded number of steps;
otherwise, the interfering thread was able to make progress.

The |push| and |remove| operations are only lock-free and therefore do not
provide an upper bound on the number of iterations. In practice however,
the average runtime is expected to be small (constant). We have verified
this experimentally, but have to refer to~\cite{Traff17:reclamationcorr,Poeter18}
for the results due to the limited space.

Finally, we claim the following.
\begin{propose}
The implementation is correct under the C++ memory model.
\end{propose}
All atomic operations are relaxed as far as possible without
sacrificing correctness with the appropriate C++ memory model
annotations. Due to the limited space we cannot show full code with
memory orderings, which we use to carefully argue that required
happens-before relationships hold as needed.  However, it is not
possible to follow the correctness arguments on the basis of the C++
memory model's semantics without the corresponding code; we have to
refer to~\cite{Poeter18} for full listings and correctness arguments
(also for the C++ implementations of the other schemes).

\lstDeleteShortInline|

\section{Experimental Comparison}
\label{sec:experiments}

We have compared Stamp-it with six other currently considered
reclamation schemes: Lock-free Reference Counting
(LFRC)~\cite{Valois:Phd}, Hazard Pointers
(HPR)~\cite{Michael:HazardPointers}, Epoch Based Reclamation
(ER)~\cite{Fraser:Phd}, New Epoch Based Reclamation
(NER)~\cite{Hart:2007:PMR:1316099.1316427}, Quiescent State Based
Reclamation (QSR)~\cite{Mckenney98read-copyupdate} and
DEBRA~\cite{Brown:2015:RML:2767386.2767436}.
All schemes have been implemented in C++ using the adapted interface
described in Section~\ref{sec:interface} and the C++ memory model. All
implementations are tuned by relaxing the atomic operations as far as
possible. Correctness arguments based on the memory model's semantics
for all implementations are provided in~\cite{Poeter18}. The full
source code is available at \url{http://github.com/mpoeter/emr}. All
results and scripts are available at \url{http://github.com/mpoeter/emr-benchmarks}.

The tests are set up similarly to Hart et
al.~\cite{Hart:2007:PMR:1316099.1316427} and we also repeat most of
those analyses. We can show here only a subset of our results, more
results can be found in~\cite{Traff17:reclamationcorr,Poeter18} and in
the appendix.

\subsection{Benchmarks}

We tested the reclamation schemes on a (1) queue, a (2) linked-list
and a (3) hash-map.  The queue is based on Michael and Scott's
design~\cite{Michael:1996:SFP:248052.248106}, the linked-list and
hash-map on Michael's improved
version~\cite{Michael:2002:HPD:564870.564881} of Harris' list-based
set~\cite{Harris:2001:PIN:645958.676105}. 
For the List benchmark the key range is twice the initial list
size. The probabilities of inserting and removing nodes are equal,
keeping the size of the list and queue data structures roughly
unchanged throughout a given run. The List benchmark has a
\emph{workload parameter} that determines the fraction of updates
(remove/insert) of the total number of operations. A workload of $0\%$
corresponds to a search-only use case, while a workload of $100\%$
corresponds to an update-only use case.

Our experiments are \emph{throughput oriented} in the following sense.
The main thread spawns $p$ child threads and starts a timer. Every
child thread performs operations on the data structure under scrutiny
until the timer expires. The parent thread calculates the average
execution time per operation by summing up the runtimes of the child
threads and their number of performed operations.  Each benchmark was
performed with 30 trials with eight seconds runtime per trial. Most of
the benchmarks focus on \emph{performance}, and calculate the
\emph{average runtime per single operation} for each trial.  Each
thread calculates its average operation runtime by dividing its
active, overall runtime by the total number of operations it
performed. The total average runtime per operation is then calculated
as the average of these per-thread runtime values.

The Queue and List benchmarks are synthetic micro-benchmarks, exactly
as by Hart et al.~\cite{Hart:2007:PMR:1316099.1316427}. The
HashMap benchmark is intended to highlight other properties of the
reclamation schemes. It 
mimics the calculation in a complex simulation where partial results
are stored in a hash-map for later reuse. These partial results are
relatively large, so in order to limit the total memory usage the
number of entries in the hash-map is kept below some threshold by
evicting old entries using a simple FIFO policy. The resulting
benchmark has the following properties:
\begin{itemize}
\item there is no upper bound on the number of nodes that are
  \emph{intentionally} blocked from reclamation.
\item the average lifetime of each \texttt{guard\_ptr} is relatively
  long.
\item the memory footprint of each node is significant.
\end{itemize}
Since there is no upper bound on the number of nodes that need to be
available for a thread, we have to use the extended hazard pointer
scheme that supports a dynamic number of hazard pointers as explained by
Michael~\cite{Michael:HazardPointers}.  The number of buckets in the
hash-map is 2048 and the maximum number of entries is 10000. There are
30000 possible partial results and every thread has to calculate or
reuse 1000 partial results per ``simulation''.  The size of a partial
result is 1024 bytes.

It is important to note that all 30 trials were performed one after
the other within the same process. This is especially important in
case of the HashMap benchmark as the hash-map is retained over the
whole runtime. This means that a result calculated in the first trial
can be found and reused in a subsequent trial. For this reason,
performance will be worse at the beginning, while the hash-map is in
the ``warm up phase'', but will improve over time when it becomes
filled and more items can be reused. But also in the other benchmarks,
it is possible that previous trials have impact on later ones, \eg due
to an already initialized memory manager. It was a deliberate design
decision to run all trials in the same process as this might more
closely reflect a real world situation.

\subsection{Environment}

\begin{figure*}[!tb]
  \centering
  \includegraphics[width=\textwidth]{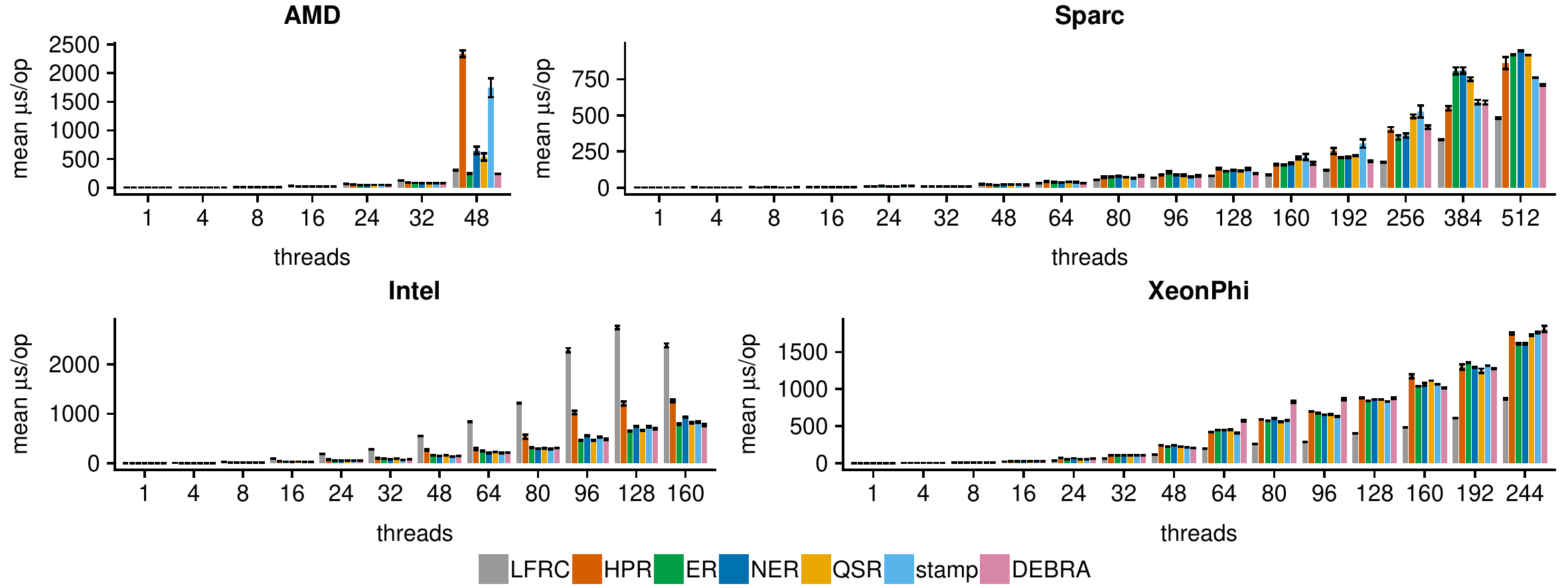}
  \caption{Queue benchmark with varying number of threads.}
  \label{fig:threads-queue}
\end{figure*}
\begin{figure*}[!tb]
  \centering
  \includegraphics[width=\textwidth]{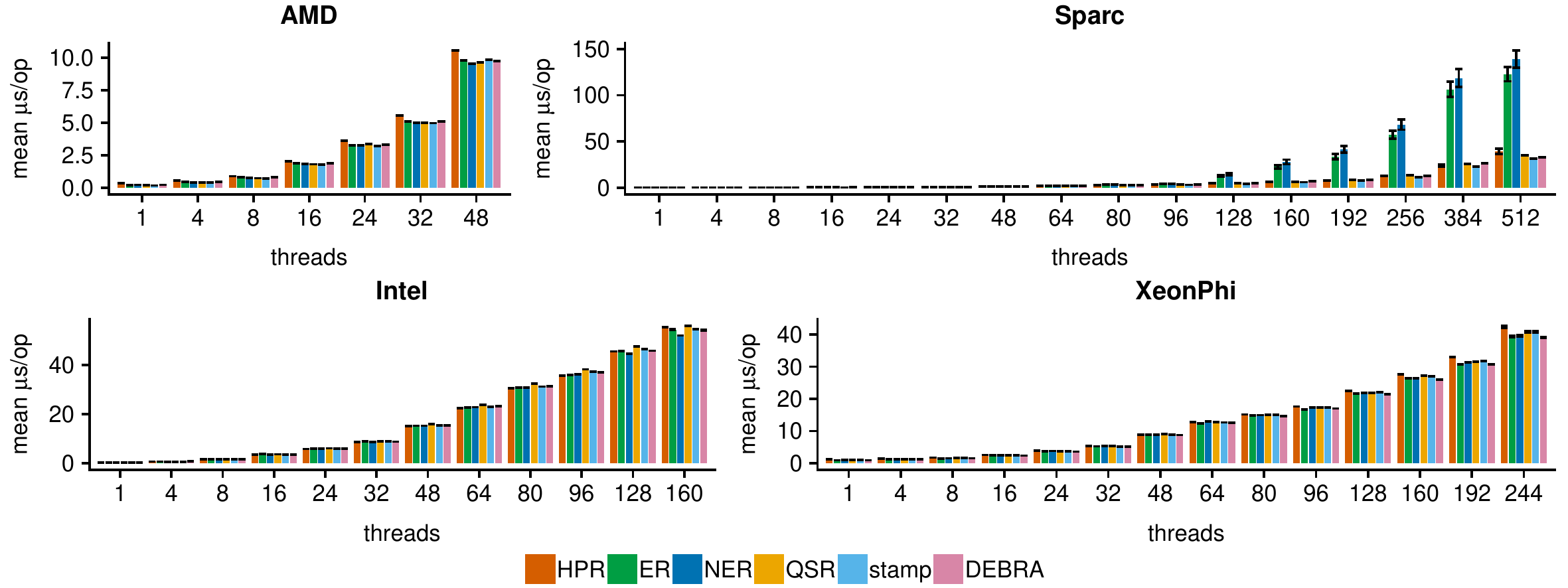}
  \caption{List benchmark with 10 elements, a workload of 20\% and varying number of threads (without LFRC).}
  \label{fig:threads-list-20}
\end{figure*}
\begin{figure*}[!tb]
  \centering
  \includegraphics[width=\textwidth]{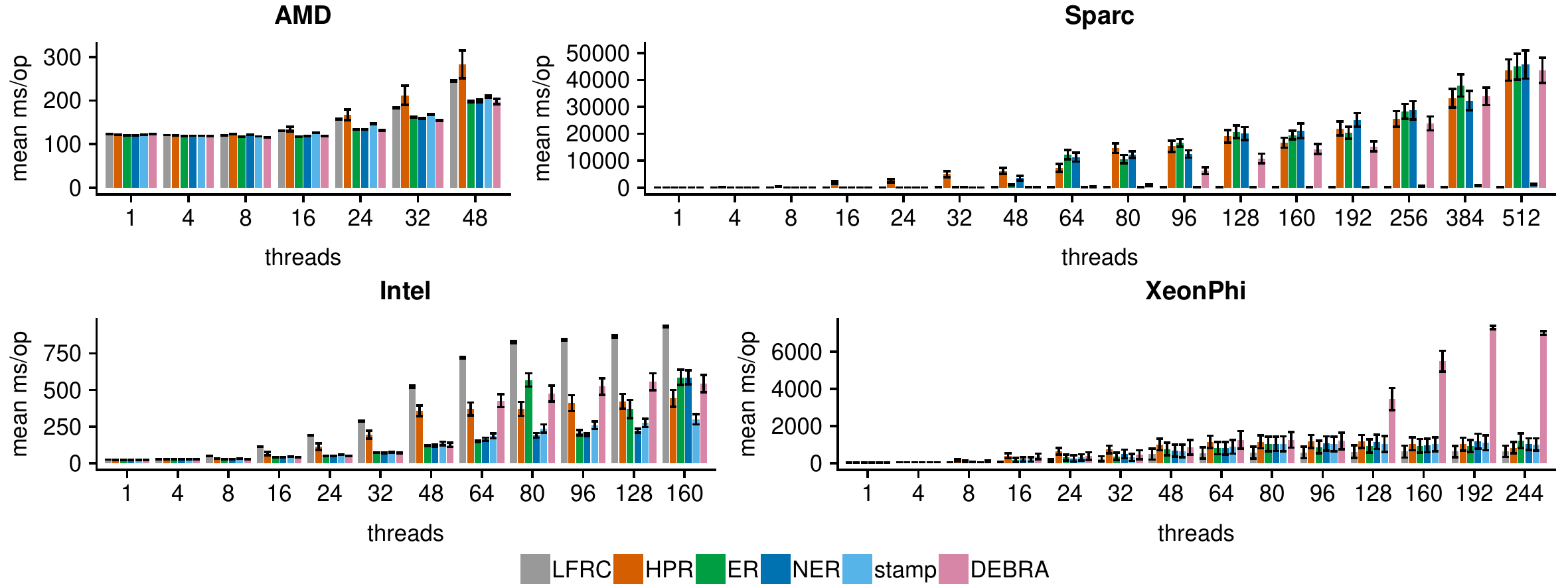}
  \caption{HashMap benchmark with varying number of threads.}
  \label{fig:threads-hash_map}
\end{figure*}

We executed our tests on four machines with different
(mi\-cro)ar\-chitectures.  Their respective characteristics are shown
in Table~\ref{tbl:Machines}. These machines all have a relatively
large number of cores and hardware supported threads, allowing us to
run our experiments at a scale not found in most prior studies.  We
did not experiment with oversubscribed cores.

\begin{longtable}{lll}
	\label{tbl:Machines} \\
	\caption{The four machines used in the experimental evaluation} \\
	\toprule
	\multirow{5}{*}{\textbf{AMD}} & CPUs & 4x AMD Opteron(tm) Processor 6168 \\
		& Frequency & max. 1.90GHz \\
		& Cores/CPU & 12 \\
		& SMT & -- \\
		& Hardware Threads & 48 \\
		& Memory & 128GB\\
		& OS & \parbox[t]{9cm}{Linux 4.7.0-1-amd64 \#1 SMP \\ Debian 4.7.6-1 (2016-10-07) x86\_64 GNU/Linux} \\
		& Compiler & gcc version 6.3.0 20170205 (Debian 6.3.0-6)\\
	\midrule
	\multirow{5}{*}{\textbf{Intel}} & CPUs &8x Intel(R) Xeon(R) CPU E7- 8850 \\
		& Frequency & max. 2.00GHz \\
		& Cores/CPU & 10 \\
		& SMT & 2x \\
		& Hardware Threads & 160 \\
		& Memory & 1TB \\
		& OS & \parbox[t]{9cm}{Linux 4.7.0-1-amd64 \#1 SMP \\ Debian 4.7.6-1 (2016-10-07) x86\_64 GNU/Linux} \\
		& Compiler & icpc version 17.0.1 (gcc version 6.0.0 compatibility)\\
	\midrule
	\multirow{5}{*}{\textbf{XeonPhi}} & CPUs & 1x Intel(R) Xeon Phi(TM) coprocessor x100 family \\
		& Frequency & max. 1.33GHz \\
		& Cores/CPU & 61 \\
		& SMT & 4x \\
		& Hardware Threads & 244 \\
		& Memory & 16GB\\
		& OS & \parbox[t]{9cm}{Linux 2.6.38.8+mpss3.8.1 \#1 SMP \\ Thu Jan 12 16:10:30 EST 2017 k1om GNU/Linux} \\
		& Compiler & icpc version 17.0.1 (gcc version 5.1.1 compatibility)\\
	\midrule
	\multirow{5}{*}{\textbf{SPARC}} & CPUs & 4x SPARC-T5-4 \\
		& Frequency & max. 3.60GHz \\
		& Cores/CPU & 16 \\
		& SMT & 8x \\
		& Hardware Threads & 512 \\
		& Memory & 1TB\\
		& OS & SunOS 5.11 11.3 sun4v sparc sun4v \\
		& Compiler & gcc version 6.3.0 (GCC) \\
	\bottomrule
\end{longtable}

We used the \texttt{jemalloc}~\cite{Evans06ascalable} memory manager on all
systems. The main reason being that on Solaris the \texttt{libc} memory manager
uses a global lock. For comparison we also ran the experiments with the
standard \texttt{libc} memory manager on all systems except SPARC; these
results are shown in Appendix~\ref{appendix:libc-results}.

ER/NER try to advance the epoch every 100 critical region entry. DEBRA
checks the next thread every 20 critical region entries. In
the List and Queue benchmarks, a \texttt{region\_guard} spans 100
benchmark operations, so this is the size of the critical region for
QSR, NER and Stamp-it. QSR executes a fuzzy barrier when it exits the
critical region. In HPR, a local retire-list is scanned once its threshold
is exceeded; the threshold is $100 + \sum_{i=0}^p K_i*2$ where $p$ is the
number of threads and $K_i$ is the number of hazard pointers for the
thread with index $i$.

\subsection{Scalability with Threads}
\label{thread-scalability}

We first study the effect of increasing the number of threads that
share a single instance of some data structure.

Figure~\ref{fig:threads-queue} shows the performance of the
reclamation schemes in the Queue benchmark. Surprisingly, LFRC
performs by far the best on Sparc and on XeonPhi, but is by far the
worst on Intel. On AMD, HPR has a huge performance drop when running
with the maximum number of threads. A similar effect can be seen by
the other schemes as well, especially NER and Stamp-it, but less
significant.  Apart from these exceptions, all schemes seem to scale
largely equally well in this scenario.

For the results of the List benchmark in
Figure~\ref{fig:threads-list-20}, LFRC has been excluded because it
performs exceedingly poor in this scenario. On AMD, Intel and XeonPhi,
all schemes are more on less on par, but on Sparc ER and NER show a
significant degradation when the number of threads grows beyond 128.
We did not investigate the reasons for this in more detail.

Finally, the results for the HashMap benchmark are shown in
Figure~\ref{fig:threads-hash_map}.  QSR has been excluded because it
scales very poorly on all architectures in this update-heavy
scenario. On AMD, ER, NER, Stamp-it and DEBRA scale almost perfectly,
while LFRC's and HPR's performance starts to degrade once the number of
threads grows beyond 16. On Intel, LFRC scales very poorly while all
other schemes scale more or less equally well, but not as well as
on AMD. On XeonPhi on the other hand, LFRC scales best while HPR's
performance starts degrading with more than 16 threads, but it again
improves with more than 128 threads. The other schemes continuously
loose performance when the number of threads grows from 16 to
${\sim}80$, but then stays more or less the same. DEBRA's performance
drastically breaks down with more than 128 threads.

\begin{figure*}[!tb]
  \centering
  \includegraphics[width=\textwidth]{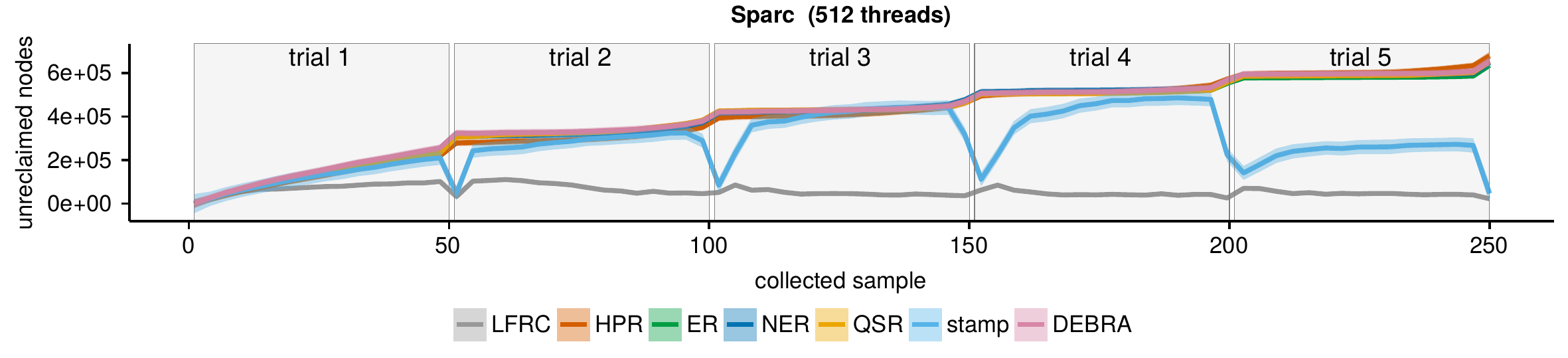}
  \caption{Number of unreclaimed nodes over time in the HashMap benchmark on Sparc. The $x$-axis is the current sample.}
  \label{fig:unreclaimed-nodes-hash_map-Sparc}
\end{figure*}
\begin{figure*}[!tb]
  \centering
  \includegraphics[width=\textwidth]{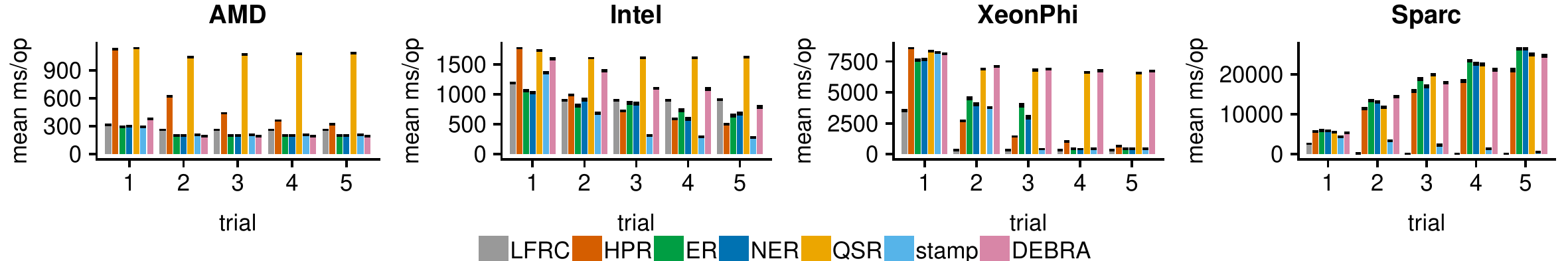}
  \caption{Development of runtime over time in the HashMap benchmark.}
  \label{fig:unreclaimed-nodes-hash_map-runtime}
\end{figure*}

The biggest surprise is the result on Sparc. Here, the performance of
HPR, ER, NER and DEBRA degrades dramatically, while LFRC and Stamp-it
scale almost perfectly. With 512 threads the performance difference
between LFRC/Stamp-it and the other schemes is a factor of ${\sim}4000$.
The reason for this will become clear when we look at the results of the
reclamation efficiency analysis in the next section.

\subsection{Reclamation Efficiency}
\label{perf-memory}

This analysis focuses on how efficiently (fast) the various schemes
actually reclaim retired nodes. An increased reclamation efficiency
can drastically reduce memory pressure, which in turn can have a
significant impact on the overall application
performance. Nonetheless, this aspect is usually disregarded in
analyses of concurrent reclamation schemes.

To measure reclamation efficiency we use thread-local performance
counters that track the number of allocated and reclaimed nodes. By
calculating the differences, we get the number of unreclaimed nodes,
which is our measurement for efficiency; a smaller number of
unreclaimed nodes means that the reclamation scheme works more
efficiently.

The plots in this analysis show the development of the number of
unreclaimed nodes over time. Each configuration is run with five
trials, each with a runtime of eight seconds. During each trial a
total of 50 samples are collected. Since the benchmarks are randomized
each configuration with the five trials is run 20 times to account for
any fluctuation in the measured samples. The plots show the smoothed
conditional means of the measured samples of those 20 runs over the
number of samples recorded during each run.

For reclamation efficiency, reference counting is the ``gold
standard''. In contrast to all other schemes there is no delay: A node
is reclaimed immediately when the last thread drops its reference to
that node. So in all the plots, LFRC can bee seen as the baseline
against which all other schemes have to be measured. One has to keep
in mind, though, that LFRC is not a general reclamation scheme, since
the reclaimed nodes cannot be returned to the memory manager, but are
stored in a global free-list.

Figure~\ref{fig:unreclaimed-nodes-hash_map-Sparc} shows the results
for the HashMap benchmark on Sparc. The number of unreclaimed nodes for
HPR, ER, NER, QSR and DEBRA is constantly increasing and does not even go
down at the end of the trials when all threads are stopped.
When a thread terminates, all schemes add the remaining nodes to a
global list. But who is responsible to reclaim them, and when? In
Stamp-it the responsibility is transferred to the ``last''
thread. Other schemes do not have a notion of a ``last'' thread, so
the global retire-list is checked by each thread when it performs
reclamation on its local retire-list. When a thread tries to reclaim
nodes from the global list it steals the whole list, reclaims all
reclaimable nodes and then re-adds the remaining nodes to the global
list.  This leads to a race during the end of a trial; whoever steals
the list might not be able to reclaim all nodes yet, but when the
remaining nodes are re-added to the global list, there might be no
threads left. Stamp-it mitigates this race as only the last thread
reclaims the global list. In addition, we can easily check
whether the global stamp has changed since reclamation has started,
so we can restart reclamation with the new stamp value. Obviously,
the effects of this race get more pronounced the more threads are
involved. 

The failure to efficiently reclaim nodes increases memory pressure,
which has a direct impact on the
runtime. Figure~\ref{fig:unreclaimed-nodes-hash_map-runtime} shows the
development of the runtime over the five trials. On Sparc we can see
that the runtime of HPR, ER, NER, QSR and DEBRA is increasing with each
trial, while with LFRC and Stamp-it it is decreasing. On the other
architectures runtime is decreasing for all schemes except QSR. This
would be the expected behavior since more results can be reused once
the hash-map has been filled.

The results for the other benchmarks and machines can be found in 
the appendix.

\section{Concluding Remarks}
\label{sec:conclusion}

This paper introduced \emph{Stamp-it}, a new, general purpose memory
reclamation scheme with attractive features. To the best of our
knowledge, this is the first non-reference counting based scheme that
does not have to scan all other threads to determine reclaimability of
a node.  We presented a large scale experimental study, comparing the
performance of Stamp-it against six other reclamation schemes on four
different architectures in various scenarios.  Our empirical results
show that Stamp-it matches or outperforms the other analyzed
reclamation schemes in almost all cases.
All of the analyzed schemes are implemented in portable, standard
conform C++, based on the standardized interface proposed by
Robison~\cite{Robison:2013}; the full source code is available at
\url{https://github.com/mpoeter/emr}.

For future work it would be interesting to look for other data
structures that could replace the doubly-linked list, \ie data
structures that have less overhead while providing all the required
properties. In this context we might also try to relax some of these
properties (e.g., use a partial order instead of a strict order for
thread entries) in order to reduce contention on the data structure.

\bibliographystyle{abbrv}
\bibliography{stampit} 

\newpage
\appendix
\section{Additional Code and Results}

\lstMakeShortInline[keywordstyle=\color{black},basicstyle=\fontsize{8}{8}\selectfont\ttfamily]|

In this appendix, we expand on some of the (mostly implementation)
details that were only briefly discussed in the main text. We also provide
additional benchmark results on the reclamation efficiency that were omitted
from the main text due to space limitations.

\subsection{Code for important methods}

The implementation of the |push| method is shown in
Listing~\ref{lst:push}.
\begin{lstlisting}[caption={The push method}, label=lst:push]
push(thread_control_block* block)
  block->next.store(head);
  marked_ptr my_prev, head_prev = head->prev.load();
  size_t stamp;
  for (;;) {
    marked_ptr head_prev2 = head->prev.load();
    if (head_prev != head_prev2) {
      head_prev = head_prev2; continue;
    }
    stamp = head->stamp.fetch_add(StampInc);
    block->stamp.store(stamp-(StampInc-PendingPush));
    if (head->prev.load() != head_prev) continue;
    my_prev = head_prev;
    block->prev.store(my_prev);
    if (head->prev.CAS(head_prev, block)) break;
  }
  block->stamp.store(stamp);
  auto link = my_prev->next.load();
  for (;;) {
    if (link.get() == block ||
        link.mark() & DeleteMark ||
        block->prev.load() != my_prev) ||
        my_prev->next.CAS(link, block))
      break;
  }
\end{lstlisting}
The implementation of the |remove| operation is shown in
Listing~\ref{lst:remove}.
\begin{lstlisting}[caption={The remove method}, label=lst:remove]
bool remove(marked_ptr block)
{
  marked_ptr prev = set_mark_flag(block->prev);
  marked_ptr next = set_mark_flag(block->next);
  bool fully_removed = remove_from_prev_list(prev, block, next);
  if (!fully_removed)
    remove_from_next_list(prev, block, next);
  auto stamp = block->stamp.load();
  block->stamp.store(stamp + NotInList);
  bool was_last = block->prev.load().get() == tail;
  if (was_last) update_tail_stamp(stamp + StampInc);
  return wasTail;
}
\end{lstlisting}
Listing~\ref{lst:remove-from-next} shows the implementation of the
|remove_from_next_list| helper function.
\begin{lstlisting}[caption={The remove\_from\_next\_list method}, label=lst:remove-from-next]
void remove_from_next_list(
  marked_ptr prev, marked_ptr removed, marked_ptr next)
{
  const auto my_stamp = removed->stamp.load();
  marked_ptr last = nullptr;
  for (;;) {
    auto next_prev = next->prev.load();
    auto next_stamp = next->stamp.load();
    if (next_prev != next->prev.load()) continue;
    if (next_stamp & (NotInList | PendingPush)) {
      if (last.get() != nullptr) {
        next = last; last.reset();
      }
      else next = next->next.load();
      continue;
    }
    auto prev_next = prev->next.load();
    auto prev_stamp = prev->stamp.load();
    if (prev_stamp > my_stamp ||
        prev_stamp & NotInList)
      return;
    if (prev_next.mark() & DeleteMark) {
      prev = prev->prev.load(); continue;
    }
    if (next.get() == prev.get()) return;
    if (remove_or_skip_marked_block(
          next, last, next_prev, next_stamp))
      continue;
    if (next_prev.get() != prev.get()) {
      move_next(next_prev, next, last); continue;
    }
    if (next_stamp <= my_stamp ||
        prev_next.get() == next.get())
      return;
    if (next->prev.load() == next_prev &&
        prev->next.CAS(prev_next, next) &&
        (next->next.load().mark() & DeleteMark) == 0)
      return;
  }
}
\end{lstlisting}
Listing~\ref{lst:mark-next} shows the implementation of the |move_next|
helper function (used in |remove_from_prev_list| and |remove_from_next_list|).
\begin{lstlisting}[caption={The mark\_next method}, label=lst:mark-next]
mark_next(marked_ptr block, size_t stamp)
{
  auto link = block->next.load();
  while (block->stamp.load() == stamp) {
    auto mark = link.mark()
    if (mark & DeleteMark ||
        block->next.compare_exchange_weak(link,
          marked_ptr(link.get(), mark | DeleteMark)))
      return true;
  }
  return false;
}
\end{lstlisting}
The implementation of the |remove_or_skip_marked_block| helper function
is shown in Listing~\ref{lst:remove-or-skip-marked-block}.
\begin{lstlisting}[caption={The remove\_or\_skip\_marked\_block method}, label=lst:remove-or-skip-marked-block]
bool remove_or_skip_marked_block(
  marked_ptr& next, marked_ptr& last, marked_ptr next_prev, stamp_t next_stamp)
{
  if (next_prev.mark() & DeleteMark) {
    if (last.get() != nullptr) {
      if (mark_next(next, next_stamp) &&
          last->prev.load() == next)
        last->prev.CAS(next, next_prev);
      next = last; last.reset();
    }
    else next = next->next.load();
    return true;
  }
  return false;
}
\end{lstlisting}
Listing~\ref{lst:update-tail} shows the implementation of the
|update_tail_stamp| method.
\begin{lstlisting}[caption={The update\_tail\_stamp method}, label=lst:update-tail]
update_tail_stamp(size_t stamp)
{
  auto last = tail->next.load();
  auto last_prev = last->prev.load();
  auto last_stamp = last->stamp.load();
  if (last_stamp > stamp &&
      last_prev.get() == tail &&
      tail->next.load() == last)
  {
    if (last.get() != head)
      stamp = last_stamp;
    else
    {
      if (stamp < last_stamp - StampInc &&
          head->prev.compare_exchange_strong(last_prev,
            make_marked(last_prev.get(), last_prev)))
        stamp = last_stamp;
    }
  }

  auto tail_stamp = tail->stamp.load();
  while (tail_stamp < stamp) {
    if (tail->stamp.compare_exchange_weak(tail_stamp, stamp))
      break;
  }
}
\end{lstlisting}

\subsection{Reclamation efficiency}
\label{appendix:reclamantion-efficiency}

\begin{figure*}
	\centering
	\includegraphics[width=\textwidth]{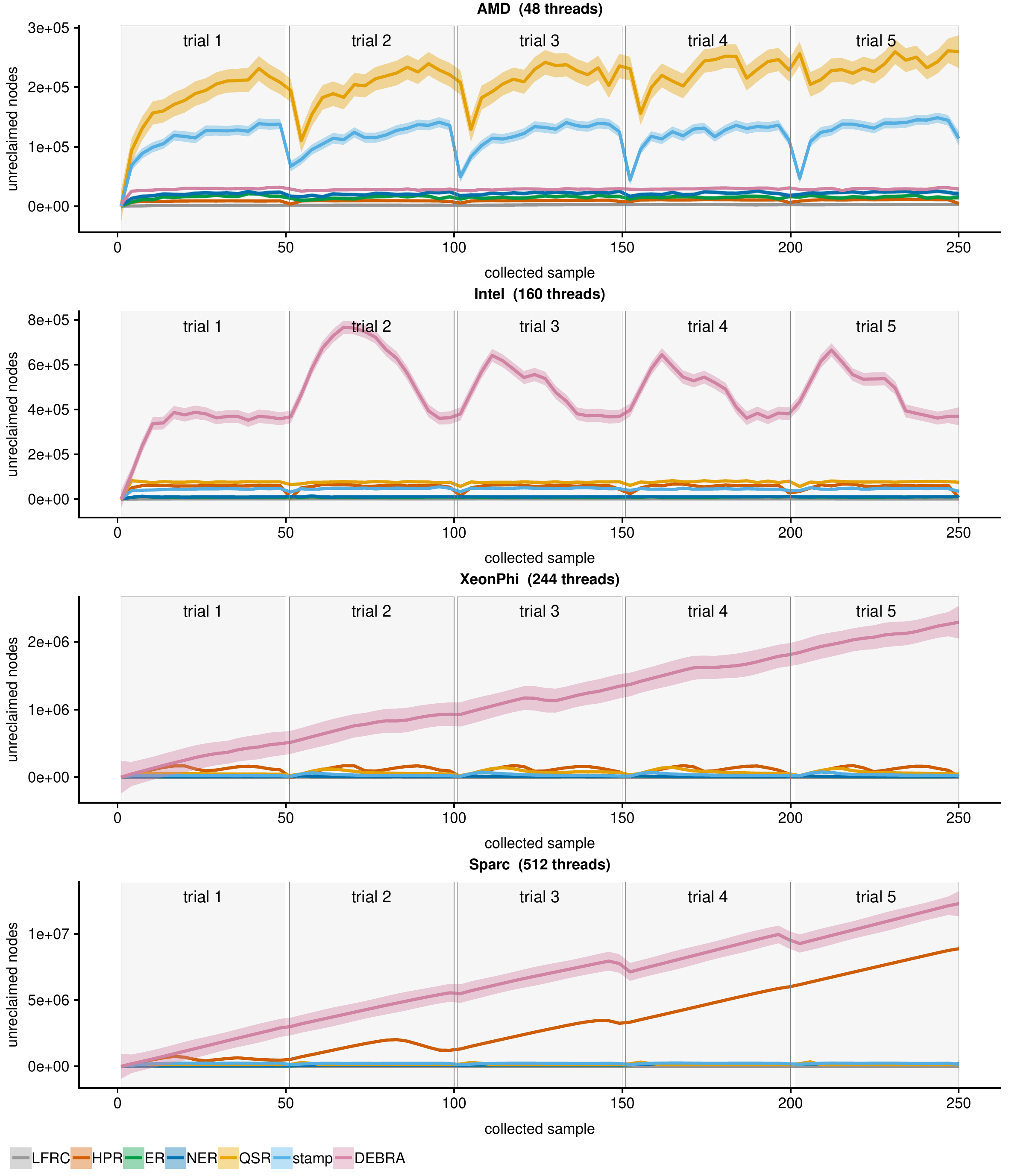}
	\caption{Number of unreclaimed of nodes over time in the Queue benchmark.}
	\label{fig:unreclaimed-nodes-queue}
\end{figure*}
\begin{figure*}
	\centering
	\includegraphics[width=\textwidth]{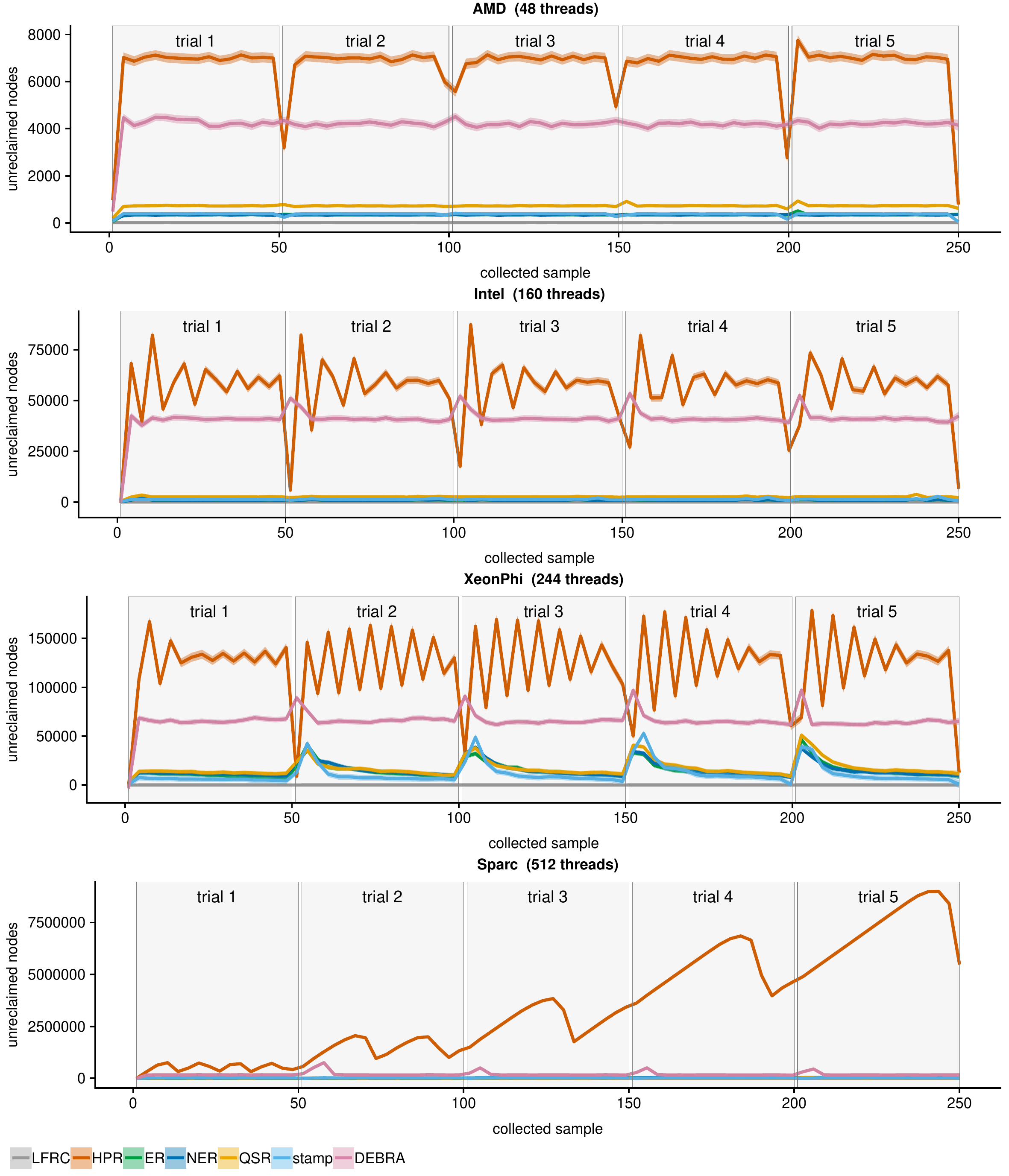}
	\caption{Number of unreclaimed nodes over time in the List benchmark with 10 elements and a workload of 20\%.}
	\label{fig:unreclaimed-nodes-list-20}
\end{figure*}
\begin{figure*}
	\centering
	\includegraphics[width=\textwidth]{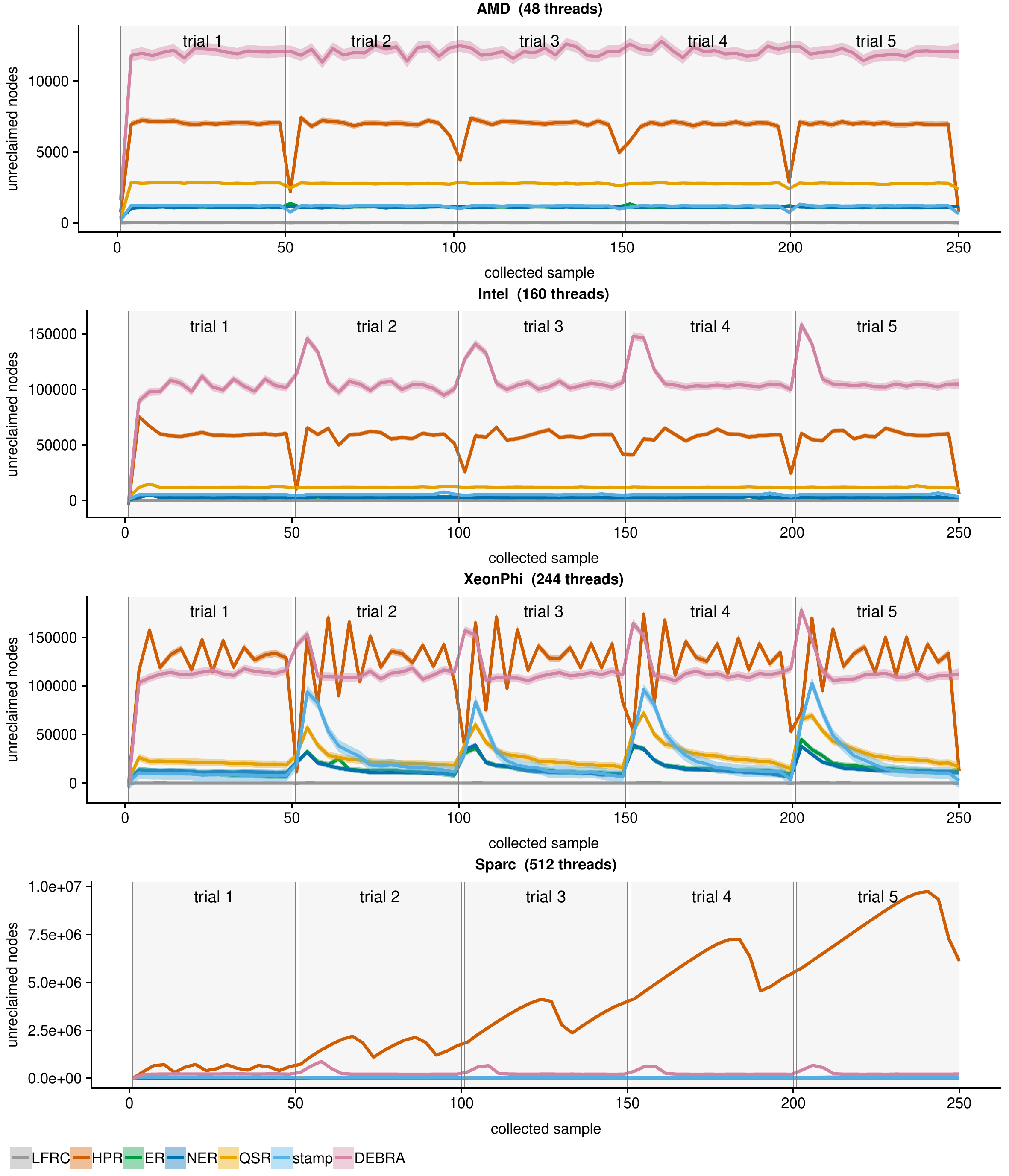}
	\caption{Number of unreclaimed nodes over time in the List benchmark with 10 elements and a workload of 80\%.}
	\label{fig:unreclaimed-nodes-list-80}
\end{figure*}
\begin{figure*}
	\centering
	\includegraphics[width=\textwidth]{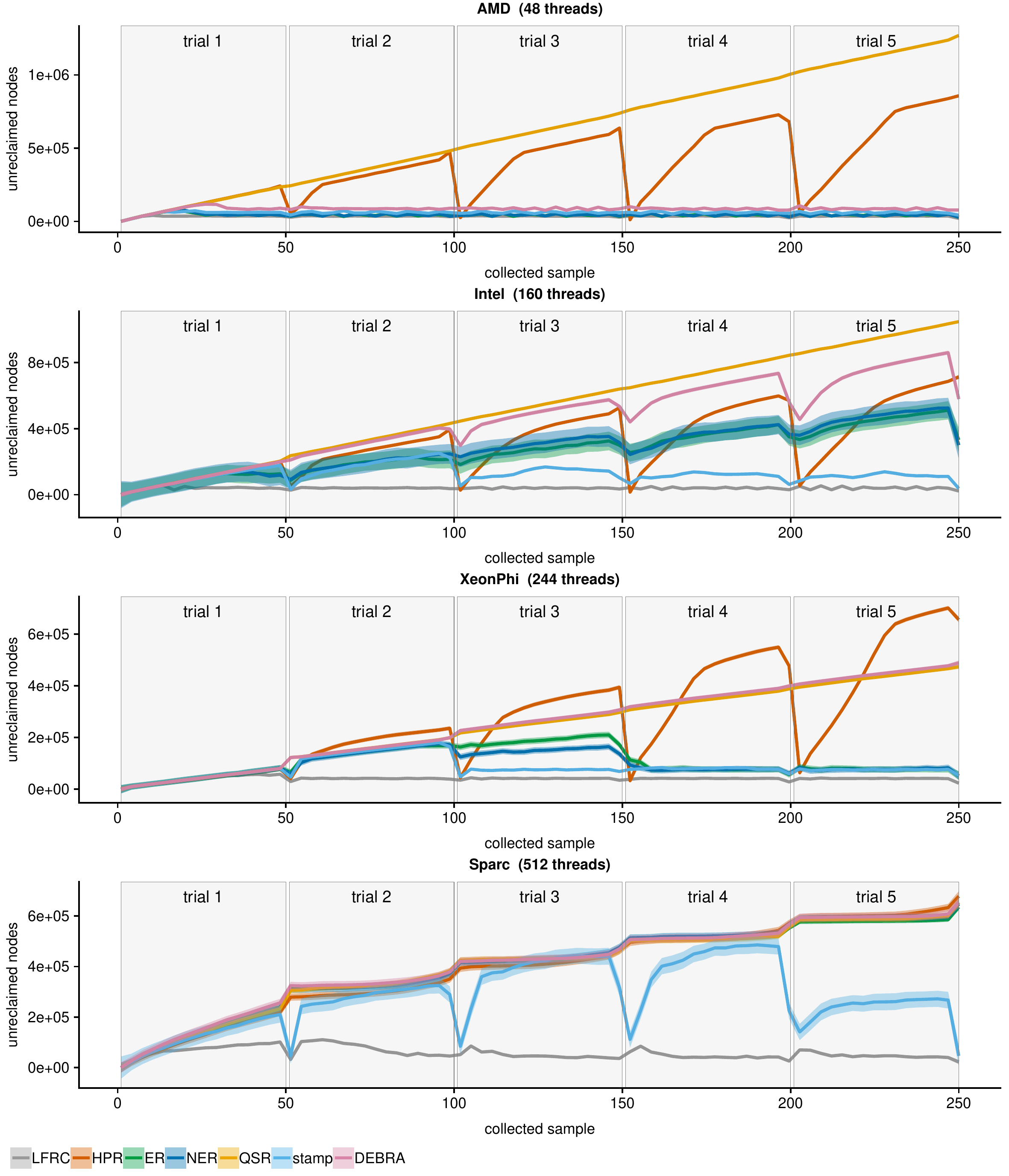}
	\caption{Number of unreclaimed nodes over time in the HashMap benchmark.}
	\label{fig:unreclaimed-nodes-hash_map}
\end{figure*}

This section contains additional results that could not be included
in the main text.
The results are shown in Figures~\ref{fig:unreclaimed-nodes-queue},
\ref{fig:unreclaimed-nodes-list-20},
\ref{fig:unreclaimed-nodes-list-80},
and~\ref{fig:unreclaimed-nodes-hash_map}. What can be seen in all
scenarios is that HPR's efficiency is inversely proportional to the
number of threads.  This is due to the threshold for the local
retire-list being calculated to achieve amortized constant time. But
this causes the number of unreclaimed nodes in the local retire-lists
to be quadratic in the number of threads. Even for the Queue benchmark
(Figure~\ref{fig:unreclaimed-nodes-queue}) and List benchmarks
(Figures~\ref{fig:unreclaimed-nodes-list-20}
and~\ref{fig:unreclaimed-nodes-list-80}) show this behavior, even
though the number of hazard pointers per thread is constant in
these scenarios. In the
HashMap benchmark (Figure~\ref{fig:unreclaimed-nodes-hash_map}) a
dynamic number of hazard pointers is used, which makes the situation
even worse.

The situation is similar for DEBRA. In order to advance the global epoch,
DEDRA does not check all $p$ threads at once, but only checks a single
thread on each critical region entry, thus distributing the costs over $p$
critical regions. But obviously with a large number of this significantly
delays the update of the global epoch, resulting in poor reclamation
efficiency.

In the Queue benchmark (see Figure~\ref{fig:unreclaimed-nodes-queue}) on
AMD QSR and Stamp-it perform relatively bad. The results for the other
architectures are dominated by the bad results  of DERBA. On Sparc, HPR
performs similarly bad as DEBRA. 

In the List benchmark (see Figures~\ref{fig:unreclaimed-nodes-list-20}
and \ref{fig:unreclaimed-nodes-list-80}) DEBRA and HPR perform
significantly worse than the other schemes on all architectures.
On Sparc, HPR performs by far the worst.

In the HashMap benchmark (Figure~\ref{fig:unreclaimed-nodes-hash_map})
we can see that QSR basically fails completely to reliably reclaim
nodes on all the architectures. The number of nodes is constantly
increasing and does not even go down at the end of the trials when all
threads are stopped. This is also the reason why QSR showed such bad
performance in the previous analysis in Section~\ref{thread-scalability}.
DEBRA performs quite good on AMD, but very poor on the other architectures.

For HPR we can also see a consistent increase in the number of
unreclaimed nodes over time, even though this number sharply drops
right at the beginning of a new trial, but also increases again very
rapidly. The only exception is Sparc, where no such drop occurs and
the number of nodes is increasing all the time. The other schemes all
perform relatively good on all architectures; the exception again
being Sparc. On Sparc HPR, ER, NER QSR and DEBRA are all performing
equally bad. The number of unreclaimed nodes is constantly increasing
and does not even go down at the end of the trials when all threads
are stopped.  This effect is probably caused by the fact that in these
schemes every thread is responsible for reclaiming its own retired
nodes. In Stamp-it we know if there is some other thread lagging
behind, so we can add nodes to a global list and let that thread take
responsibility for reclaiming them. This allows Stamp-it to more
reliably reclaim nodes, especially at the end of each trial.


\subsection{Results for \texttt{libc}}
\label{appendix:libc-results}
This section contains the same results as shown in
Section~\ref{sec:experiments} and
Appendix~\ref{appendix:reclamantion-efficiency}, but using the
standard \texttt{libc} memory manager on AMD, Intel and XeonPhi. On
Sparc we still used \texttt{jemalloc} since the \texttt{libc} memory
manager on Solaris uses a global lock.

The results do not show significant differences. The overall performance
is somewhat lower compared to the \texttt{jemalloc} results, especially on
Intel, but the distribution of the measured runtime/operation is very similar
for all schemes, in all experiments and on all machines, i.e., the impact of
the memory manager is equally big/small for all schemes.

\begin{figure*}[!tb]
  \centering
  \includegraphics[width=\textwidth]{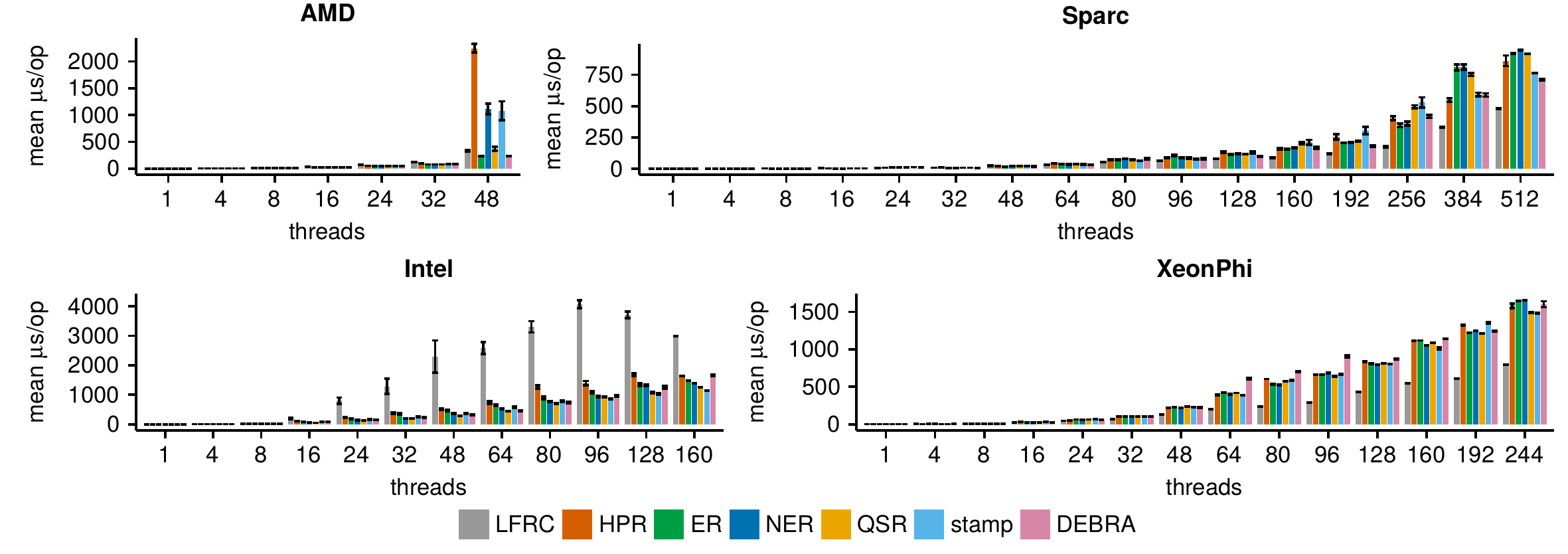}
  \caption{Queue benchmark with varying number of threads.}
  \label{fig:threads-queue-libc}
\end{figure*}
\begin{figure*}[!tb]
  \centering
  \includegraphics[width=\textwidth]{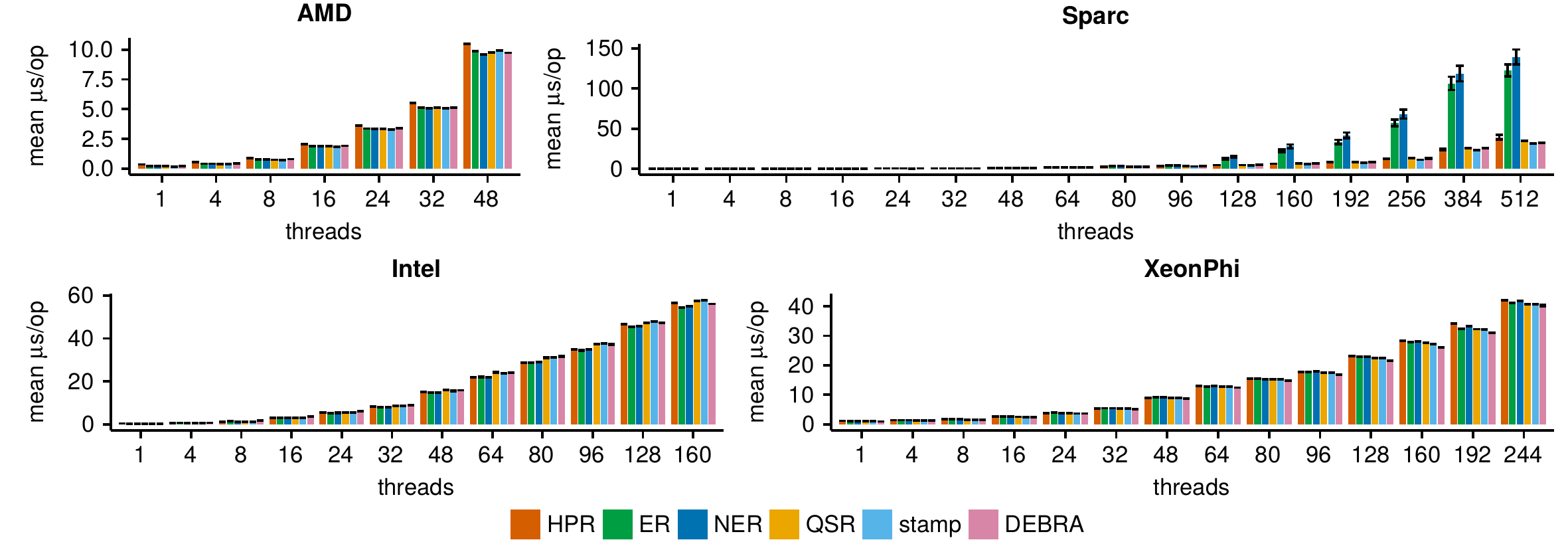}
  \caption{List benchmark with 10 elements, a workload of 20\% and varying number of threads (without LFRC).}
  \label{fig:threads-list-20-libc}
\end{figure*}
\begin{figure*}[!tb]
  \centering
  \includegraphics[width=\textwidth]{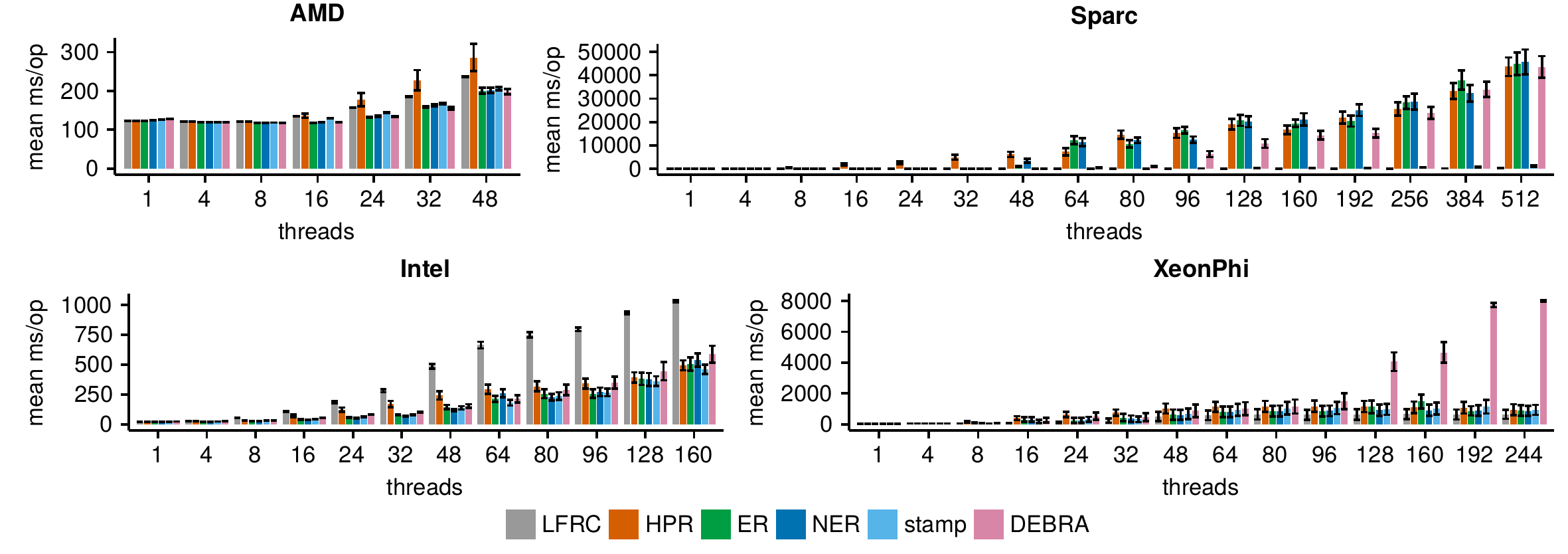}
  \caption{HashMap benchmark with varying number of threads.}
  \label{fig:threads-hash_map-libc}
\end{figure*}

\begin{figure*}[!tb]
  \centering
  \includegraphics[width=\textwidth]{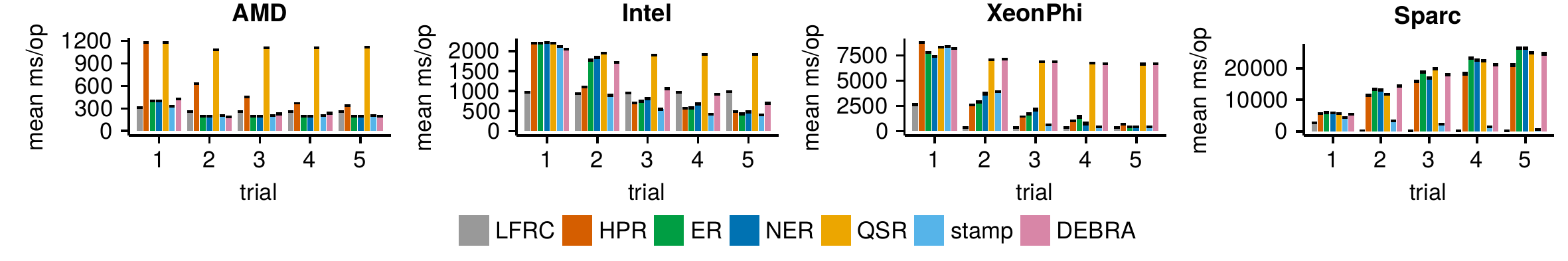}
  \caption{Development of runtime over time in the HashMap benchmark.}
  \label{fig:unreclaimed-nodes-hash_map-runtime-libc}
\end{figure*}

\begin{figure*}
	\centering
	\includegraphics[width=\textwidth]{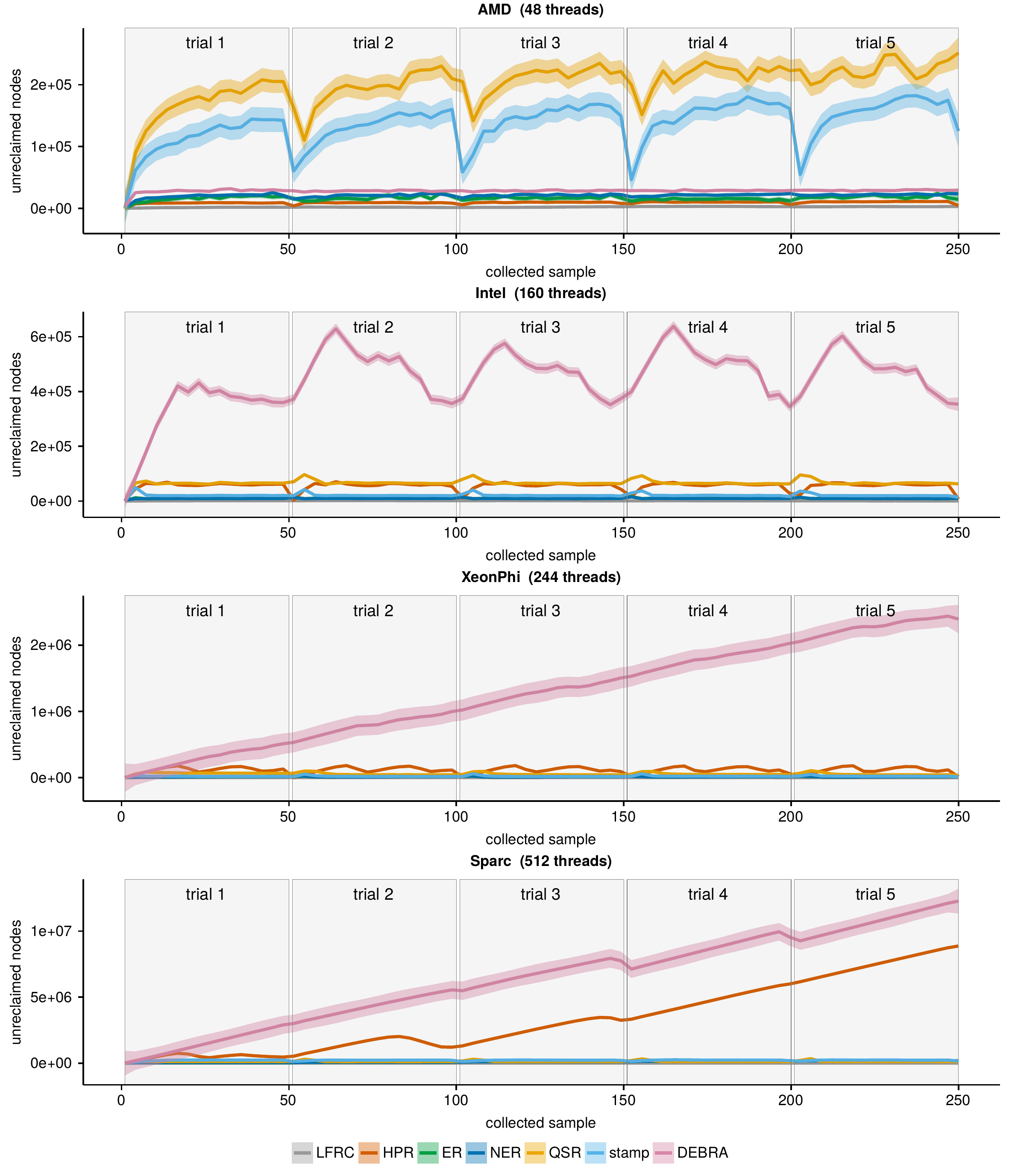}
	\caption{Number of unreclaimed of nodes over time in the Queue benchmark.}
	\label{fig:unreclaimed-nodes-queue-libc}
\end{figure*}
\begin{figure*}
	\centering
	\includegraphics[width=\textwidth]{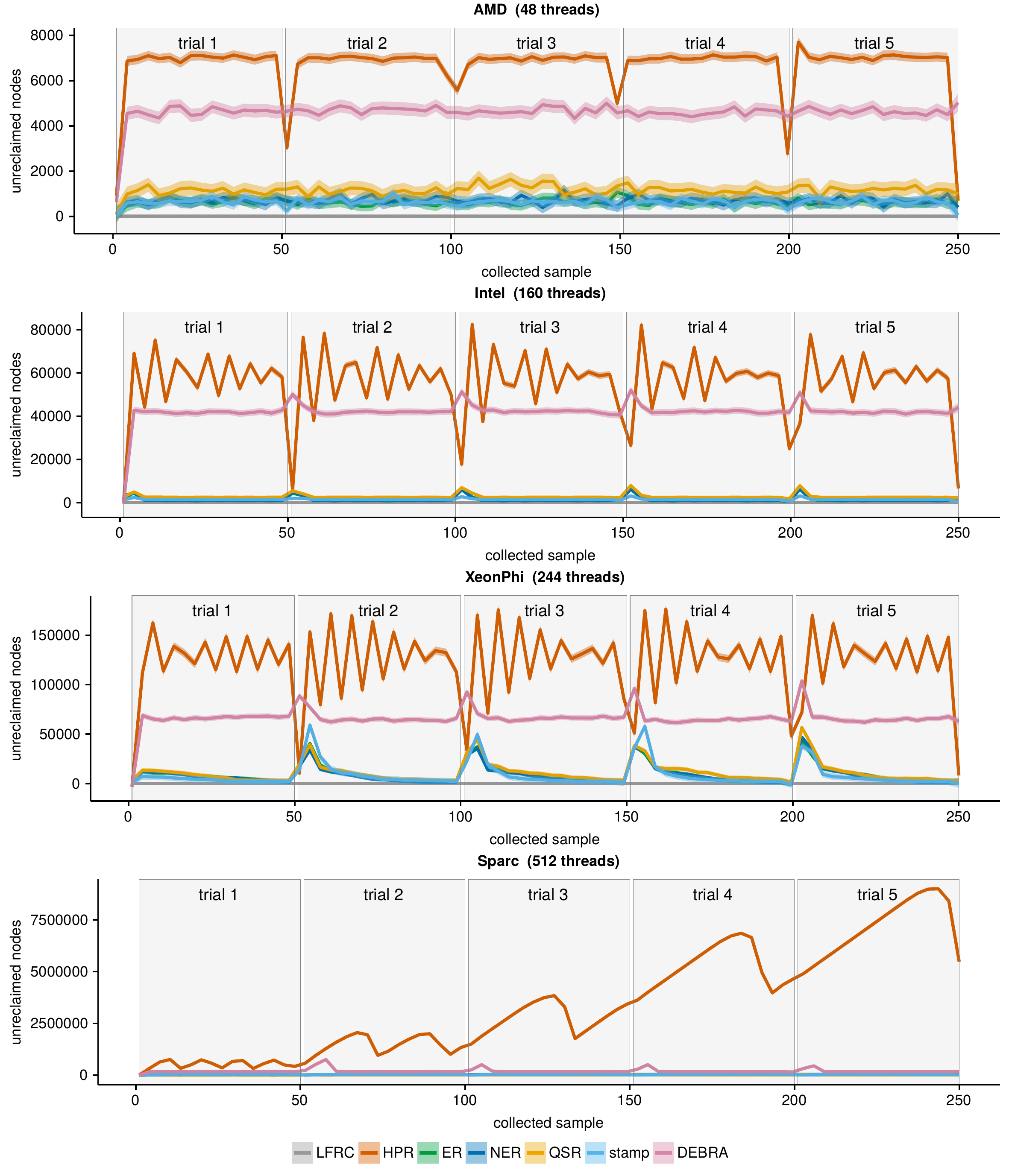}
	\caption{Number of unreclaimed nodes over time in the List benchmark with 10 elements and a workload of 20\%.}
	\label{fig:unreclaimed-nodes-list-20-libc}
\end{figure*}
\begin{figure*}
	\centering
	\includegraphics[width=\textwidth]{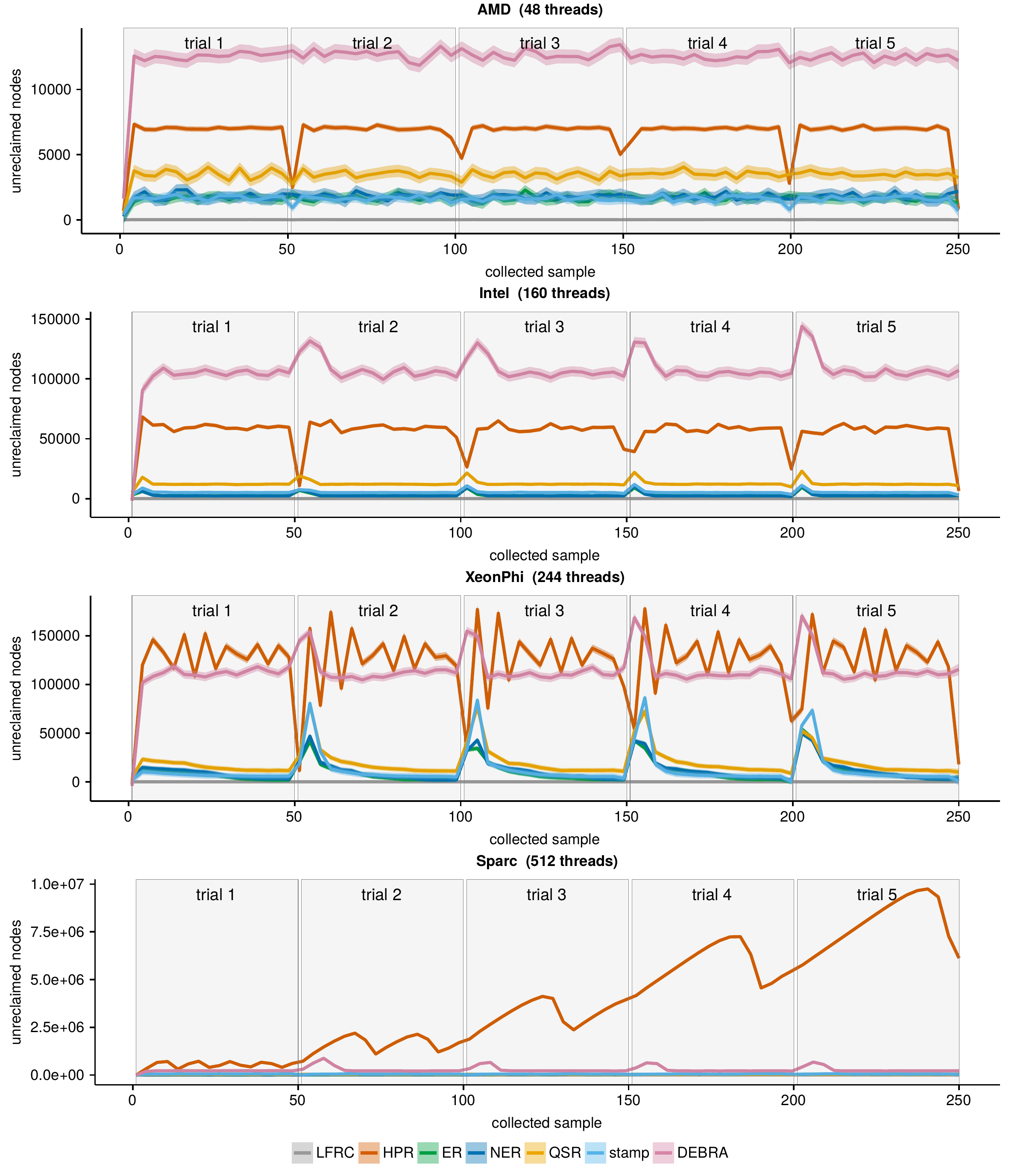}
	\caption{Number of unreclaimed nodes over time in the List benchmark with 10 elements and a workload of 80\%.}
	\label{fig:unreclaimed-nodes-list-80-libc}
\end{figure*}
\begin{figure*}
	\centering
	\includegraphics[width=\textwidth]{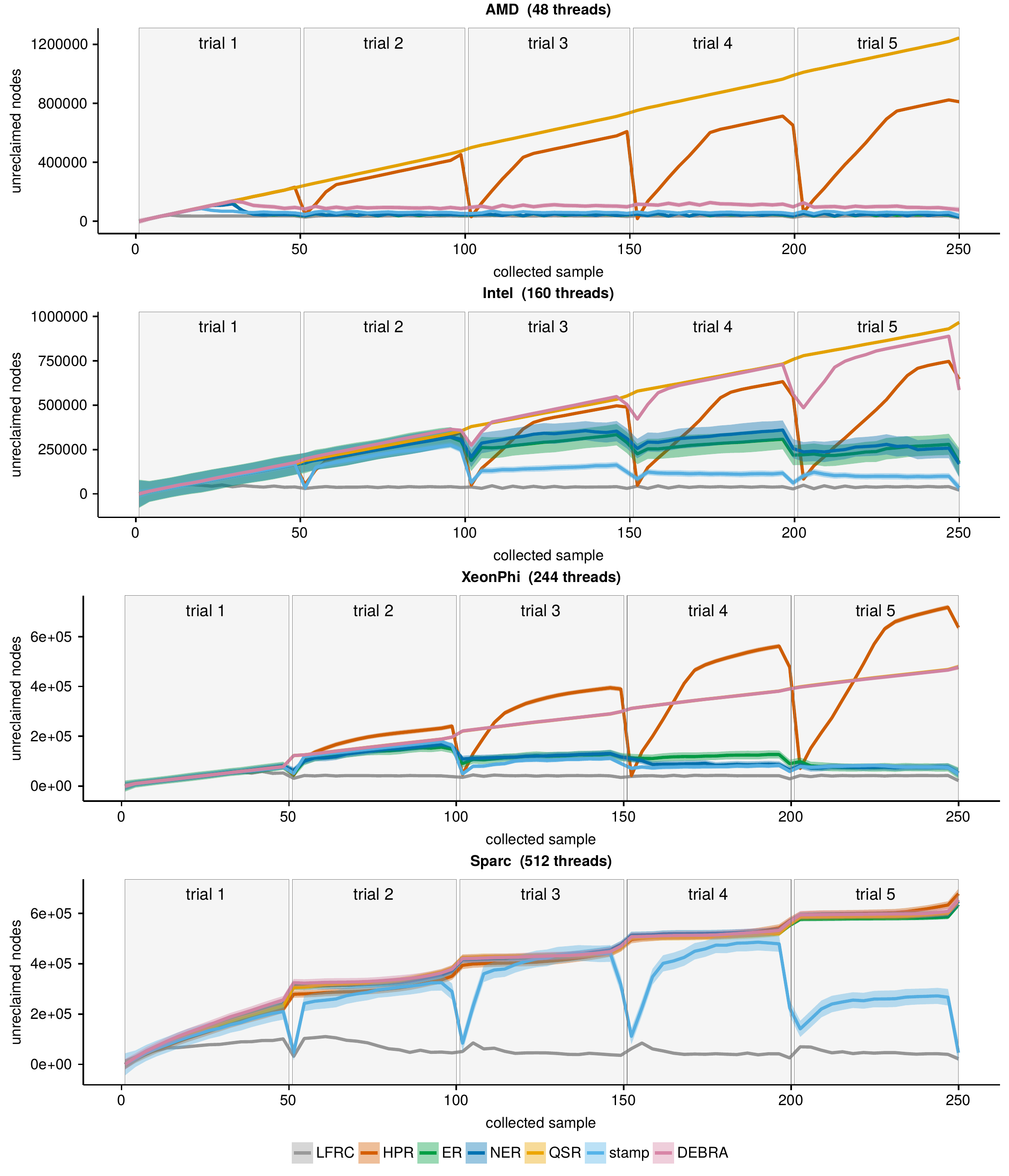}
	\caption{Number of unreclaimed nodes over time in the HashMap benchmark.}
	\label{fig:unreclaimed-nodes-hash_map-libc}
\end{figure*}

\end{document}